\documentclass{IEEEtran}
\usepackage{cite}
\usepackage{amsmath,amssymb,amsfonts}
\usepackage{algorithmic}
\usepackage{graphicx}
\usepackage{textcomp}
\def\BibTeX{{\rm B\kern-.05em{\sc i\kern-.025em b}\kern-.08em
    T\kern-.1667em\lower.7ex\hbox{E}\kern-.125emX}}

\makeatletter
\renewcommand\@seccntformat[1]{%
  \csname the#1\endcsname\quad 
}
\makeatother

\usepackage{dcolumn}
\usepackage{bm}
\usepackage{xcolor}
\usepackage{float}
\usepackage{upgreek}

\newcommand{\uve}[1]{\hat{\boldsymbol{#1}}}
\newcommand{\ve}[1]{\mathbf{#1}}

\newcommand{\te}[1]{\overline{\overline{\ve{#1}}}}
\newcommand{\ves}[1]{\boldsymbol{#1}}
\newcommand{\tes}[1]{\overline{\overline{\ves{#1}}}}

\newcommand{\pati}[1]{
}

\begin{document}
\title{Electromagnetic Theory of Metasurface \\ Perfect Magnetic Conductor (PMC)}
\author{Oscar Céspedes Vicente, Karim Achouri, \IEEEmembership{Member, IEEE}, and Christophe Caloz, \IEEEmembership{Fellow, IEEE}
}

\maketitle

\begin{abstract}
Artificial magnetic conductors (AMCs) mimic the idealized boundary condition of a perfect magnetic conductor (PMC), which reflects electromagnetic waves with a preserved electric field and inverted magnetic field. Despite their usefulness, existing AMC implementations often rely on complex or impractical designs, and lack a clear electromagnetic theory explaining their behavior, especially under oblique or polarization-diverse incidence. This work addresses these limitations by presenting a rigorous electromagnetic framework for PMC metasurfaces based on dipolar and quadrupolar surface susceptibilities within the generalized sheet transition conditions (GSTCs) formalism. We show that achieving polarization- and angle-independent PMC behavior requires a specific set of heteroanisotropic (nonlocal) susceptibilities, and we derive closed-form expressions for angular scattering that include higher-order multipole contributions. A physically realizable, asymmetric metasurface structure is then designed to satisfy these theoretical conditions. Despite its geometric asymmetry, the proposed structure exhibits a isotropic PMC response at resonance, confirmed by full-wave simulations and multipolar susceptibility extraction. These results demonstrate how properly engineered surface multipoles can yield angularly independent magnetic boundary conditions using only thin, passive metallic layers. This work bridges the gap between AMC design and electromagnetic theory, and enables a new class of angle-independent metasurface reflectors for more accurate simulations, optimizations and innovative AMC designs.
\end{abstract}



\section{Introduction}

\pati{Background}
Maxwell’s equations’ symmetry in the presence of both electric \emph{and} magnetic charges and currents theoretically suggests the existence of magnetic monopoles~\cite{jackson2021classical}. However, despite Dirac’s argument based on charge quantization~\cite{dirac1931quantised} and subsequent studies~\cite{tHooft1974monopole,Polyakov1974monopole,Witten1979dyon}, magnetic monopoles have never been experimentally observed. Nevertheless, the concept of hypothetical magnetic charges and currents inspired the development of \emph{artificial magnetic conductors (AMCs)}---engineered surfaces that emulate the boundary conditions of an ideal, \emph{perfect magnetic conductors (PMCs)}. Seminal realizations of such AMCs include the mushroom-type AMC of Sievenpiper {\it et al.}~\cite{sievenpiper1999high} and the complementary cross-potent AMC of Ma {\it et al.}~\cite{ma1998realisation,yang1999novel}, introduced in the late 1990s, which showed promise for high-gain planar antenna~\cite{feresidis2005artificial} and transverse electromagnetic rectangular waveguide applications~\cite{yang1999novel}. Despite persistent challenges, AMCs remain foundational components in modern electromagnetics and continue to drive the search for simpler, more versatile and theoretically grounded implementations. These limitations mainly concern undesired polarization and angular dependence, and most mitigation strategies have proven impractical. For instance, Monorchio {\it et al.} proposed AMCs that rely on extreme dielectric parameters~\cite{monorchio2006optimal}, whereas Hashemi {\it et al.} employed a complex geometry of oppositely handed spiral resonators connected by vias~\cite{hashemi2013dual}. On the modeling side, related approaches have been restricted to lumped-element circuit analogs~\cite{shahparnia2004simple,costa2010analysis,padooru2011circuit} or periodic transverse-resonant transmission-line networks~\cite{rahman2001transmission,luukkonen2008simple,rodriguez2015analytical}. However, these models lack a rigorous electromagnetic foundation and often fail to capture essential physical behaviors such as polarization dependence, angular dispersion and advanced effects like gyrotropy and nonreciprocity. These limitations highlight a critical gap: the absence of a comprehensive electromagnetic theory for AMC surfaces, along with the need for structurally simple and physically realizable implementations consistent with such a theoretical framework.

\pati{Contribution}
This work addresses the above gap through two key contributions. First, it introduces a rigorous \emph{electromagnetic theory} for AMCs, formulated in terms of \emph{multipolar surface susceptibilities} within the framework of \emph{generalized sheet transition conditions (GSTCs)}. Second, it presents a physically realizable \emph{PMC metasurface structure} derived from this theory that exhibits an elevation-angle- and polarization-independent response.

\pati{Organization}
The remainder of the paper is organized as follows. Section~\ref{sec:magnetic_conductor} reviews the operation and fundamental properties of magnetic conductors, establishing their phenomenological equivalence to magnetic surfaces. Section~\ref{sec:multipole_theory} introduces the theoretical framework based on dipole--quadrupole generalized sheet transition conditions (DQ-GSTC), formulates the angular scattering response and derives the susceptibility conditions required for perfect magnetic conductor (PMC) behavior. Section~\ref{sec:structure} presents the proposed metasurface structure and explains how it achieves quasi-azimuthal independence. Section~\ref{sec:demonstration} provides full-wave simulation results that validate the theoretical predictions and demonstrate the structure’s practical performance. Finally, conclusions are presented in Sec.~\ref{sec:concl}.

\section{Magnetic Conductor}\label{sec:magnetic_conductor}

\begin{figure}[H]
        \centering
        \includegraphics[width=1.05\columnwidth]{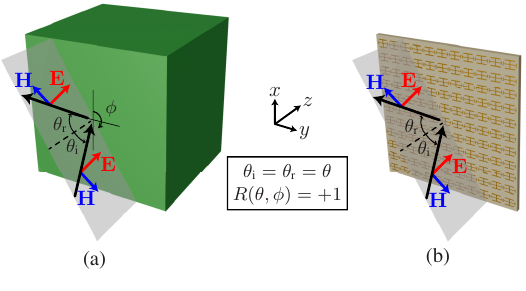}
        \vspace{-11mm}\caption{Perfect magnetic conductor (PMC) under oblique plane-wave incidence, with incidence and reflection elevation angles, $\theta_\text{i}$ and $\theta_\text{r}$, and arbitrary polarization, in the arbitrary incidence plane $\phi$. (a)~Hypothetical bulk material. (b)~Equivalent metasurface.}
        \label{fig:mag_cond_op}
    \end{figure}

\pati{Local Introduction}
A PMC is an ideal conductor that reflects all incident electromagnetic fields with the electric field preserved and the magnetic field inverted. Figure~\ref{fig:mag_cond_op} illustrates the canonical PMC configuration: a plane wave impinges obliquely in an arbitrary azimuthal plane $\phi$, with equal incidence and reflection elevation angles $\theta_\text{i} = \theta_\text{r} = \theta$, i.e., \emph{specular reflection}. As shown in Fig.~\ref{fig:mag_cond_op}(a), the PMC can be conceptualized as a bulk magnetic material; however, since no field exists in perfect conductor, practical implementations are typically realized using \emph{metasurfaces}, as in Fig.~\ref{fig:mag_cond_op}(b), emulating the PMC behavior via engineered surface magnetic currents.

\pati{Hypothetical Magnetic Conductor}
A hypothetical PMC, illustrated in Fig.~\ref{fig:mag_cond_op}(a), can be modeled as a bulk magnetic medium characterized by boundary conditions that are the electromagnetic dual of those for a perfect electric conductor (PEC). These PMC boundary conditions are expressed as
\begin{subequations}
        \begin{align}
        &\boldsymbol{\hat{z}}\times \Delta\mathbf{H}=0,\label{eq:PMC_Je0}\\
        &\boldsymbol{\hat{z}}\times\Delta\mathbf{E}=-\mathbf{J}_\text{m0},\label{eq:PMC_Jm0}\\
        &\boldsymbol{\hat{z}}\cdot\Delta\mathbf{D}=0,\label{eq:PMC_rhoe0}\\
        &\boldsymbol{\hat{z}}\cdot\Delta\mathbf{B}=\rho_\text{m0},\label{eq:PMC_rhom0}
        \end{align}
where $\Delta\ves{\psi}=\ves{\psi}(z=0^+)-\ves{\psi}(z=0^-)$ ($\ves{\psi}=\ve{E},\ve{H}$) denotes the discontinuity across the PMC interface. The surface quantities $\ve{J}_\text{m0}$ and $\rho_\text{m0}$ satisfy the magnetic current continuity condition~\cite{rothwell2018electromagnetics}
\begin{equation}\label{eq:PMC_cont_curr}
    \nabla_\parallel\cdot \ve{J}_\text{m0}+\frac{\partial\rho_\text{m0}}{\partial t}=0.
\end{equation}
\end{subequations}
In the far field, the scattering response of a PMC is fully determined by Eq.~\eqref{eq:PMC_Je0} and the free-space Maxwell equations. Specifically, the condition $\Delta \ve{H}_\parallel = 0$ implies that the tangential magnetic field is perfectly reflected, $\ve{H}_{\text{r},\parallel} = -\ve{H}_{\text{i},\parallel}$, while the tangential electric field is preserved, $\ve{E}_{\text{r},\parallel} = \ve{E}_{\text{i},\parallel}$. As a result, the reflection coefficients are $R_\text{s} = +1$ for s polarization (s-pol) and $R_\text{p} = +1$ for p polarization (p-pol), for all frequencies, $f$, elevation angles, $\theta$, and azimuthal incidence plane angles, $\phi$\footnote{%
This follows from the phase-matching and specular-reflection relations 
$\mathbf{k}_{\text{r},\parallel} = \mathbf{k}_{\text{i},\parallel}$ and 
$k_{\text{r},z} = -k_{\text{i},z}$, together with the tangential projection 
of Ampère-Maxwell’s law in free space,
$\mathbf{E}_\parallel = 
\hat{\mathbf{z}}\times\hat{\mathbf{z}}\times[(\mathbf{k}/k)\times\eta_0\mathbf{H}$],
and magnetic Gauss’s law,
$H_z=-\mathbf{k}_\parallel\cdot\mathbf{H}_\parallel/k_z$~\cite{jackson2021classical}.
These relations yield the tangential electric field of the incident or reflected wave:
\[
\ve{E}_{\text{i/r},\parallel} =
-\frac{\ve{k}_{\text{i/r},\parallel}\cdot\eta_0\ve{H}_{\text{i/r},\parallel}}
{k_{\text{i/r},z}}\frac{\hat{\ve{z}}\times\ve{k}_{\text{i/r},\parallel}}{k}
-\frac{k_{\text{i/r},z}}{k}\hat{\ve{z}}\times\eta_0\ve{H}_{\text{i/r},\parallel},
\]
where the identity 
$\ve{a}\times(\ve{b}\times\ve{c})=(\ve{a}\cdot\ve{c})\ve{b}-(\ve{a}\cdot\ve{b})\ve{c}$ has been used.}.
Finally, since the PMC is a perfect conductor, its fields are entirely supported by surface magnetic currents and charges confined to an infinitesimally thin skin layer at the interface.

\pati{Artificial PMC Metasurface}
As previously mentioned, AMCs are typically implemented as engineered surfaces—metasurfaces—that emulate ideal PMC boundary conditions through effective surface magnetic currents~\cite{achouri2021electromagnetic}. As illustrated in Fig.~\ref{fig:mag_cond_op}(b), these metasurfaces provide a phenomenologically equivalent alternative to hypothetical volumetric magnetic materials, offering practical advantages in fabrication and integration. Although true magnetic charges and currents do not exist in nature, artificial surface magnetism in AMCs arises from the collective response of oscillating electric charges, giving rise to effective macroscopic magnetic surface current densities. As will be shown, this artificial electromagnetic behavior is fundamentally produced by multipole moments of the underlying electric current distribution~\cite{raab2005multipole}. Moreover, to faithfully replicate PMC reflection properties in the far field, the metasurface must reproduce the correct equivalent magnetic response independently of both the incidence elevation $\theta$ and azimuth $\phi$ and for any polarization, thereby achieving fully angular- and polarization-independent PMC operation.

\section{Multipoles Theory}\label{sec:multipole_theory}

\pati{DQ-GSTCs}
To capture the angular scattering response of metasurfaces, we adopt a multipolar framework that extends conventional dipolar generalized sheet\footnote{The assumed deeply subwavelength thickness of a metasurface eliminates Fabry--Perot resonances, allowing the structure to be accurately described by \emph{surface} polarization densities within a \emph{sheet} model formalism.} transition conditions (D-GSTCs)~\cite{achouri2021electromagnetic} to include quadrupolar effects, forming the dipole--quadrupole GSTCs (DQ-GSTCs)~\cite{achouri2022multipolar}
\begin{subequations}\label{eq:DQ-GSTC}
\begin{align}
&\uve{z}\times\Delta\mathbf{H}=\ve{J}_\text{e0}^\text{pol},\label{eq:DQ-GSTC_Je0}\\
&\uve{z}\times\Delta\mathbf{E}=-\ve{J}_\text{m0}^\text{pol},\label{eq:DQ-GSTC_Jm0}\\
&\uve{z}\cdot\mathbf{D}=\rho_\text{e0}^\text{pol},\\
&\uve{z}\cdot\mathbf{B}=\rho_\text{m0}^\text{pol},
\end{align}
where $\Delta\ves{\psi}$ denotes the difference of the fields at both side of the metasurface and where the surface polarization currents and charges are
\begin{align}
\begin{split}\label{eq:Jepol}
\ve{J}_\text{e0}^\text{pol}&=j\omega\ve{P}_{\parallel}-\uve{z}\times\nabla_{\parallel} M_z\\
&+\frac{k^2}{2}\boldsymbol{\hat{z}}\times\left(\te{S}\cdot\boldsymbol{\hat{z}}\right)
-j\frac{\omega}{2}\nabla_\parallel\cdot\left(\overline{\overline{\mathbf{Q}}}-Q_{zz}\overline{\overline{\mathbf{I}}}\right)_\parallel\\
&+\frac{1}{2}\uve{z}\times\nabla_{\parallel}\left(\left(\nabla_{\parallel}\otimes\boldsymbol{\hat{z}}+\uve{z}\otimes\nabla_{\parallel}\right)\cdot\te{S}\right),
\end{split}\\
\begin{split}\label{eq:Jmpol}
\ve{J}_\text{m0}^\text{pol}&=j\omega\mu_0\ve{M}_{\parallel}+\frac{1}{\epsilon_0}\uve{z}\times\nabla_{\parallel} P_z\\
&-\frac{k^2}{2\epsilon_0}\boldsymbol{\hat{z}}\times\left(\te{Q}\cdot\boldsymbol{\hat{z}}\right)-j\frac{\omega\mu_0}{2}\nabla_\parallel\cdot\left(\overline{\overline{\mathbf{S}}}-S_{zz}\overline{\overline{\mathbf{I}}}\right)_\parallel\\
&-\frac{1}{2\epsilon_0}\uve{z}\times\nabla_{\parallel}\left(\left(\nabla_{\parallel}\otimes\boldsymbol{\hat{z}}+\uve{z}\otimes\nabla_{\parallel}\right)\cdot\te{Q}\right),
\end{split}\\
\begin{split}\label{eq:rhoepol}
        \rho_{\text{e0}}^\text{pol}&=\\
        &\hspace{-1cm}-\nabla_\parallel\cdot\left(\ve{P}_{\parallel}-j\frac{\omega\epsilon_0\mu_0}{2}\uve{z}\times\left(\te{S}\cdot\uve{z}\right)_\parallel-\frac{1}{2}\nabla_\parallel\cdot\left(\te{Q}-Q_{zz}\te{I}\right)_\parallel\right),
        \end{split}\\
        \begin{split}\label{eq:rhompol}
        \rho_{\text{m0}}^\text{pol}&=\\
        &\hspace{-1cm}-\nabla_\parallel\cdot\mu_0\left(\ve{M}_{\parallel}+j\frac{\omega}{2}\uve{z}\times\left(\te{Q}\cdot\uve{z}\right)_\parallel-\frac{1}{2}\nabla_\parallel\cdot\left(\te{S}-S_{zz}\te{I}\right)_\parallel\right),
        \end{split}
\end{align}
\end{subequations}
where $\ve{P}$ and $\ve{M}$ denote the electric and magnetic dipole moment densities, and $\te{Q}$ and $\te{S}$ the electric and magnetic quadrupole moment densities, the subscript $\parallel$ denotes the parallel part and $z$ the perpendicular part with respect to the metasurface plane, $\otimes$ denotes the dyadic outer product. 

\pati{DQ-GSTC Susceptibilities}
The multipolar polarization densities in Eq.~\eqref{eq:DQ-GSTC} are related to the fields and their derivatives as
\begin{subequations}\label{eq:multipole_susceptibility_based}
\begin{equation}
\begin{pmatrix} P_i \\M_i \\ Q_{il}\\ S_{il}\end{pmatrix}=\tes{\upchi}\cdot
\begin{pmatrix} E_{\text{av},j} \\ H_{\text{av},j} \\ \partial_k E_{\text{av},j}\\ \partial_k H_{\text{av},j}\end{pmatrix},
\end{equation}
where $\ves{\psi}_\text{av}=\big(\ves{\psi}(z=0^+)+\ves{\psi}(z=0^-)\big)/2$ denotes the field averages at both side of the metasurface, by the susceptibility block tensor~\cite{achouri2023spatial}
\begin{equation}\label{eq:susceptibility_compact_form}
\hspace{-5mm}\tes{\upchi}=
\begin{pmatrix}
\epsilon_0 \chi_{\text{ee},ij} & \frac{1}{c} \chi_{\text{em},ij} & \frac{\epsilon_0}{2k} \chi'_{\text{ee},ikj} & \frac{1}{2ck} \chi'_{\text{em},ikj}\\
\frac{1}{\eta_0} \chi_{\text{me},ij} & \chi_{\text{mm},ij} & \frac{1}{2\eta_0k} \chi'_{\text{me},ikj} & \frac{1}{2k} \chi'_{\text{mm},ikj}\\
\frac{\epsilon_0}{k} Q_{\text{ee},ilj} & \frac{1}{ck} Q_{\text{em},ilj} & \frac{\epsilon_0}{2k^2} Q'_{\text{ee},ilkj} & \frac{1}{2ck^2} Q'_{\text{em},ilkj}\\
\frac{1}{\eta_0 k} S_{\text{me},ilj} & \frac{1}{k} S_{\text{mm},ilj} & \frac{1}{2\eta_0 k^2} S'_{\text{me},ilkj} & \frac{1}{2k^2} S'_{\text{mm},ilkj}
\end{pmatrix},
\end{equation}
\end{subequations}
where each component is a tensor whose rank is determined by its number of indices, with each index ranging from 1 to 3 to represent the three spatial dimensions. This results in a total of $4\times3^2+8\times3^3+4\times3^4=576$ susceptibilities components.
$\chi_{\text{ee},ij}$, $\chi_{\text{em},ij}$, $\chi_{\text{me},ij}$ and $\chi_{\text{mm},ij}$ are the zeroth-order (local) dipolar susceptibilities of the conventional D-GSTC bianisotropic tensors. They respectively describe the electric response to an electric field, electric response to a magnetic field, magnetic response to an electric field and magnetic response to a magnetic field. These tensors govern the direct and cross-coupling between the tangential fields and the induced electric and magnetic dipole moments.
$\chi_{\text{ab},ikj}'$ ($\text{ab}=\{\text{ee}, \text{em}, \text{me}, \text{mm}\}$) are first-order dipolar susceptibilities associated with the spatial derivative $\partial_k\ves{\psi}_{\text{av},j}$ of the incident fields. These (nonlocal) tensors characterize spatial dispersion effects and describe how dipole moments are induced not only by the fields but also by their spatial gradients.
$Q_{\text{ab},ilj}$ and $S_{\text{ab},ilj}$ are zeroth-order quadrupolar susceptibilities that relate the averaged fields $\ves{\psi}_{\text{av},j}$ to the induced electric ($\te{Q}$) and magnetic ($\te{S}$) quadrupole moments. These third-rank tensors extend the response beyond dipolar approximations and account for field-induced quadrupolar distributions in the metasurface.
$Q'_{\text{ab},iljk}$ and $S'_{\text{ab},iljk}$ are first-order quadrupolar susceptibilities associated with the field derivatives $\partial_k\ves{\psi}_{\text{av},j}$. These fourth-rank tensors model higher-order spatial-dispersion effects and essentially capture nonlocal phenomena in highly spatially dispersive metasurfaces. As we shall see later, many of all these susceptibilities are inter-dependent or physically indistinguishable, leading to a natural grouping into independent degrees of freedom, which dramatically reduces the number of independent parameters~\cite{achouri2022multipolar}.

\pati{GSTC System Description}
The angular reflection and transmission coefficients of a general\footnote{A general metasurface is penetrable and therefore supports transmission when illuminated from either side.} metasurface are fully determined by the surface polarization currents in Eqs.~\eqref{eq:DQ-GSTC_Je0} and~\eqref{eq:DQ-GSTC_Jm0}. These equations provide four equations per polarization, two for wave illumination in the positive-$z$---forward---direction of incidence and two for the negative-$z$---backward---direction of incidence, for four unknowns, the forward and backward reflection and transmission coefficients.

\pati{Resolution of the System}
We start with the case of plane-wave illumination in the $xz$ ($\phi = 0$) plane, considering the s-pol and p-pol separately. The first step is to specify the electric and magnetic field components relevant to each polarization (see Appendix~\ref{appendix:Fields_Specifications}), expressed in terms of the reflection and transmission coefficients, $R^{\pm}(\theta)$ and $T^{\pm}(\theta)$, where $\pm$ refers to forward ($+\hat{\bm{z}}$) or backward ($-\hat{\bm{z}}$) propagation. These fields are substituted into the DQ-GSTC to relate field discontinuities and averages to the polarization multipole densities. To simplify the system, we assume the metasurface is \emph{nongyrotropic}~\cite{lavigne2018susceptibility}, which decouples the set of four DQ-GSTC equations into two independent equations per polarization and per illumination direction. The symmetry of the metasurface with respect to the plane $(xy)-z$ reduces then the number of independent susceptibility components (see Appendix~\ref{appendix:Susceptibility_Tensor}), and the remaining interdependent susceptibilities are re-expressed in terms of a minimal set of dipolar and quadrupolar contributions, capturing the dominant multipolar behavior. Finally, to ensure a well-posed extraction of the scattering parameters, fourth-order corrections---specifically ${S'}_\text{mm}^{xzxz}$ for s-pol and ${Q'}_\text{ee}^{xzxz}$ for p-pol---are neglected (see Appendix~\ref{appendix:Well-Posed}) and the resulting reduced DQ-GSTC are finally solved to yield the angular scattering parameters for s-pol (see Appendix~\ref{appendix:Scatt_Para})\footnote{Only the s-pol case is shown for brevity.}.

\begin{figure}[H]
    \centering
    \includegraphics[width=1\columnwidth]{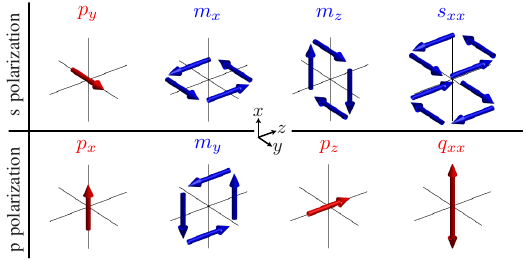}
    \vspace{-6mm}\caption{Representation of the multipole moments of a metasurface characterized by DQ-GSTC [Eq.~\eqref{eq:DQ-GSTC}] in terms of infinitesimal currents (arrows) for plane-wave incidence in the $xz$ ($\phi=0$ in Fig.~\ref{fig:mag_cond_op}) plane and for both s- and p-pol.}
	\label{fig:multipoles_terms_current}
\end{figure}

\pati{Angular Scattering}
The general expressions for the s-pol scattering parameters is (see Appendix~\ref{appendix:Scatt_Para})
        \begin{subequations}\label{eq:spol_ang_scattering}
            \begin{equation}\label{eq:spol_ang_scattering_R}
                \begin{aligned}
                R^{\pm}_s&(\theta)=\Big[-2jk\chi_\text{ee}^{yy}\mp2jk\left(\chi_\text{em}^{yx}-\chi_\text{me}^{xy}\right)c_\theta\\
                &2jk\chi_\text{mm}^{xx}s_\theta^2-2jk\chi_\text{mm}^{zz}c_\theta^2\pm2k\left({\chi'}_\text{mm}^{xxz}-{\chi'}_\text{mm}^{zxx}\right)c_\theta s_\theta^2\\
                &+2jk{S'}_\text{mm}^{xxxx}c_\theta^2 s_\theta^2\Big]/\Delta
                \end{aligned}
            \end{equation}
            and
            \begin{equation}\label{eq:spol_ang_scattering_T}
            \begin{aligned}
                T^\pm_s&(\theta)=\Big[k^2\chi_\text{ee}^{yy}\chi_\text{mm}^{xx}+\left(2\mp jk\chi_\text{em}^{yx}\right)\left(2\mp jk\chi_\text{me}^{xy}\right)\\
                &+k^2\Big(\chi_\text{ee}^{yy}{S'}_\text{mm}^{xxxx}+\chi_\text{mm}^{xx}\chi_\text{mm}^{zz}+j\chi_\text{em}^{yx}{\chi'}_\text{mm}^{xxz}+j\chi_\text{me}^{xy}{\chi'}_\text{mm}^{zxx}\\
                &\mp2\left({\chi'}_\text{mm}^{xxz}+{\chi'}_\text{mm}^{zxx}\right)\Big)s_\theta^2+\\
                &+k^2\left(\chi_\text{mm}^{zz}{S'}_\text{mm}^{xxxx}+{\chi'}_\text{mm}^{xxz}{\chi'}_\text{mm}^{zxx}\right)s_\theta^4\Big]c_\theta/\Delta,
            \end{aligned}
            \end{equation}
            with
            \begin{equation}
            \begin{aligned}
                &\Delta=2jk\chi_\text{ee}^{yy}+\left(4+k^2\chi_\text{em}^{xy}\chi_\text{me}^{xy}-k^2\chi_\text{ee}^{yy}\chi_\text{mm}^{xx}\right)c_\theta\\
                &+2jk\chi_\text{mm}^{xx}c^2_\theta+2jk\chi_\text{mm}^{zz}s^2_\theta\\
                &-k^2\left(\chi_\text{mm}^{xx}\chi_\text{mm}^{zz}+\chi_\text{ee}^{yy}{S'}_\text{mm}^{xxxx}+j\chi_\text{em}^{yx}{\chi'}_\text{mm}^{xxz}+j\chi_\text{me}^{xy}{\chi'}_\text{mm}^{zxx}\right)c_\theta s^2_\theta\\
                &+2jk{S'}_\text{mm}^{xxxx}c^2_\theta s^2_\theta-k^2\left(\chi_\text{mm}^{zz}{S'}_\text{mm}^{xxxx}+{\chi'}_\text{mm}^{xxz}{\chi'}_\text{mm}^{zxx}\right)c_\theta s^4_\theta,
            \end{aligned}
            \end{equation}
            \end{subequations}
            where the top (resp. bottom) sign is associated to forward (resp. backward) illumination, and $c_\theta=\cos(\theta)$ and $s_\theta=\sin(\theta)$.
The expressions in Eqs.~\eqref{eq:spol_ang_scattering_R} and~\eqref{eq:spol_ang_scattering_T} are rational functions of the incidence angle $\theta$ and explicitly reveal how the various multipole susceptibilities shape the angular response. Figure~\ref{fig:multipoles_terms_current} illustrates the dipolar and quadrupolar current distributions associated with the susceptibilities appearing in these expressions.

\pati{PMC Susceptibility Conditions}
Angle-independent PMC scattering coefficients are obtained by nullifying all the angle-dependent terms in Eq.~\eqref{eq:spol_ang_scattering}\footnote{The angular dependence in Eqs.~\eqref{eq:spol_ang_scattering} is nontrivial because these relations have a rational structure in which trigonometric functions and susceptibilities are nonlinearly coupled, making the extraction of angle-invariant conditions difficult. In contrast, the DQ-GSTC parameter formulation (see Appendix~\ref{appendix:PMC_susc}) provides a more tractable framework. Owing to the linear dependence of the DQ-GSTCs on the trigonometric terms ($\cos\theta$, $\sin\theta$), the angle-dependent contributions can be directly identified and canceled, thereby facilitating the derivation of angularly independent DQ-GSTC conditions.}, which yields (see Appendix~\ref{appendix:PMC_susc})
    \begin{subequations}\label{eq:spol_RT_ang_ind}
    \begin{align}
        R_\text{s}^{\pm}(\theta)&=\pm j2\frac{k(\chi_\text{em}^{yx}-\chi_\text{me}^{xy})}{k^2\chi_\text{em}^{yx}\chi_\text{me}^{xy}+4}\label{eq:Rspm_ang_ind},\\
        T_\text{s}^{\pm}(\theta)&=\frac{(2\pm jk\chi_\text{em}^{yx})(2\pm jk\chi_\text{me}^{xy})}{k^2\chi_\text{em}^{yx}\chi_\text{me}^{xy}+4}.
    \end{align}
    \end{subequations}
Note that these relations involve \emph{only heteroanisotropic susceptibilities}, since all the homoanisotropic susceptibilities exhibited angle-dependent coefficients and were therefore removed. Assuming \emph{reciprocity}, $\chi_\text{me}^{xy} = -\chi_\text{em}^{yx}$~\cite{achouri2021electromagnetic}, and enforcing the PMC condition, $R_\text{s}^{+} = +1$, yields then $\chi_\text{em}^{yx} = -j2/k$, resulting in $R_\text{s}^{\pm} = \pm1$ and $T_\text{s}^{\pm} = 0$, which corresponds to PMC and PEC responses in the forward and backward direction, respectively. The full set of susceptibility conditions for s-pol and p-pol in the $xz$ incidence plane are summarized in Tab.~\ref{tab:mag_group_susc}. They correspond to a \emph{heteroanisotropic, nonlocal\footnote{Bianisotropy is a weak form of nonlocality~\cite{caloz2020chiralVolII}.}, omega-type~\cite{achouri2021electromagnetic} metasurface}, with tensors
    \begin{equation}\label{eq:PMC_heteroanisotropic}
        \tes{\upchi}_\text{em}=
        \begin{pmatrix}
            0 & j\frac{2}{k} \\
        -j\frac{2}{k} & 0
        \end{pmatrix}\quad \text{and}\quad \tes{\upchi}_\text{me}=
        \begin{pmatrix}
            0 & j\frac{2}{k} \\
            -j\frac{2}{k} & 0
        \end{pmatrix}.
    \end{equation}
\begin{table}[H]
    \centering
   	\caption{PMC susceptibility conditions for s-pol and p-pol in the $xz$ ($\phi=0$) incidence plane.}
   	\label{tab:mag_group_susc}
        \setlength{\tabcolsep}{35 pt}
        \renewcommand*{\arraystretch}{1.6}
        \begin{tabular}{ll}
    		\hline \multicolumn{1}{c}{s polarization\textbf{}} & \multicolumn{1}{c}{p polarization}\\ \hline
            $\chi_\text{ee}^{yy}=0$ & $\chi_\text{ee}^{xx}=0$ \\ \hline
    		$\chi_\text{em}^{yx}=-j2/k$ & $\chi_\text{em}^{xy}=j2/k$\\ \hline
    		$\chi_\text{mm}^{xx}=0$ & $\chi_\text{mm}^{yy}=0$\\ \hline
    		$\chi_\text{me}^{xy}=j2/k$ & $\chi_\text{me}^{yx}=-j2/k$\\ \hline
    		$\chi_\text{mm}^{zz}=0$ & ${Q'}_\text{ee}^{xxxx}=0$\\ \hline
    		${\chi'}_\text{mm}^{zxx}=0$ & ${\chi'}_\text{ee}^{xxz}=0$\\ \hline
    		${\chi'}_\text{mm}^{xxz}=0$ & ${\chi'}_\text{ee}^{zxx}=0$\\ \hline
    		${S'}_\text{mm}^{xxxx}=0$ & $\chi_\text{ee}^{zz}=0$\\ \hline
    	\end{tabular}
    \end{table}

\pati{Forward and Backward Behaviors}
The multipolar response corresponding to the PMC solution in Eqs.~\eqref{eq:PMC_heteroanisotropic} is obtained by inserting these relations into Eq.~\eqref{eq:multipole_susceptibility_based}. For forward illumination, the induced-dipole responses are
\begin{subequations}\label{eq:dipolar_pred_f}
    \begin{equation}
        \ve{P}^{+}_\parallel=\frac{1}{c}\tes{\upchi}_\text{em}\ve{H}_{\text{av},\parallel}^{+}=\ve{0}
    \end{equation}
    and
    \begin{equation}
        \ve{M}^{+}_\parallel=\frac{1}{\eta_0}\tes{\upchi}_\text{me}\ve{E}_{\text{av},\parallel}^{+}=-j\frac{1}{k}\frac{2E_0}{\eta_0}\begin{pmatrix}
            -1\\
            \cos(\theta)
        \end{pmatrix},
    \end{equation}
    \end{subequations}
while, for backward illumination (see Appendix~\ref{appendix:Fields_Specifications}), they are
\begin{subequations}\label{eq:dipolar_pred_b}
    \begin{equation}
        \ve{P}^{-}_\parallel=\frac{1}{c}\tes{\upchi}_\text{em}\ve{H}_{\text{av},\parallel}^{-}=-j\frac{1}{k}\epsilon_0 2E_0\begin{pmatrix}
            1\\
            \cos(\theta)
        \end{pmatrix}
    \end{equation}
    and
    \begin{equation}
        \ve{M}^{-}_\parallel=\frac{1}{\eta_0}\tes{\upchi}_\text{me}\ve{E}_{\text{av},\parallel}^{-}=\ve{0}.
    \end{equation}
    \end{subequations}
The forward relations in Eqs.~\eqref{eq:dipolar_pred_f} correspond to the surface currents
\begin{subequations}\label{eq:dipolar_pred_cur}
    \begin{align}
        \ve{J}_{\text{e}0,\parallel}^{\text{pol}+}&=j\omega\ve{P}_\parallel^{+}=\ve{0},\\
        \ve{J}_{\text{m}0,\parallel}^{\text{pol}+}&=j\omega\mu_0\ve{M}_\parallel^{+}=2E_0\begin{pmatrix}
            -1\\
            \cos(\theta)
        \end{pmatrix}
    \end{align}
    \end{subequations}
for the incident field $\ve{E}_\text{i} = E_0(\cos(\theta), 1, -\sin(\theta))^T$, which satisfies the PMC requirement in Eqs.~\eqref{eq:PMC_Je0} and~\eqref{eq:PMC_Jm0}, and specifies the related magnetic current.

\pati{Planes $\phi=\pi/2$}
The results in Tab.~\ref{tab:mag_group_susc} and Eqs.~\eqref{eq:dipolar_pred_f}--\eqref{eq:dipolar_pred_cur} were derived for incidence in the $xz$ plane ($\phi=0$). The corresponding expressions for incidence in the $yz$ plane ($\phi=\pi/2$) can in fact be obtained simply by exchanging $x$ and $y$, as done in Tab.~\ref{tab:mag_group_susc_phi_pi2}. This might appear inconsistent with a $\phi=\pi/2$ rotational permutation, where $x\rightarrow y$ and $y\rightarrow-x$, yet the apparent discrepancy is resolved by noting that the susceptibilities are defined as \emph{ratios} of polarization densities to fields, according to Eq.~\eqref{eq:multipole_susceptibility_based}. The sign changes induced by the plane rotation cancel out between numerator and denominator for the homo- (electric--electric, magnetic--magnetic) terms, but must be accounted for in the off-diagonal hetero- (electric--magnetic, magnetic--electric) terms, since these represent cross-terms of the fields. Moreover, given the subwavelength nature of the metasurface unit cell, we assume that the response for incidence at $\phi=\pi/4$ (corresponding to a $\sqrt{2}$-larger projected dimension) is essentially a superposition of the $\phi=0$ and $\phi=\pi/2$ cases. Azimuthal symmetry is therefore automatically ensured once the $\phi=\pi/2$-rotation symmetry is satisfied.

\begin{table}[H]
    \centering
   	\caption{PMC susceptibility conditions for s-pol and p-pol in the $xz$ ($\phi=\pi/2$) incidence plane.}
   	\label{tab:mag_group_susc_phi_pi2}
        \setlength{\tabcolsep}{35 pt}
        \renewcommand*{\arraystretch}{1.6}
        \begin{tabular}{ll}
    		\hline \multicolumn{1}{c}{s polarization\textbf{}} & \multicolumn{1}{c}{p polarization}\\ \hline
            $\chi_\text{ee}^{xx}=0$ & $\chi_\text{ee}^{yy}=0$ \\ \hline
    		$\chi_\text{em}^{xy}=j2/k$ & $\chi_\text{em}^{yx}=-j2/k$\\ \hline
    		$\chi_\text{mm}^{yy}=0$ & $\chi_\text{mm}^{xx}=0$\\ \hline
    		$\chi_\text{me}^{yx}=-j2/k$ & $\chi_\text{me}^{xy}=j2/k$\\ \hline
    		$\chi_\text{mm}^{zz}=0$ & ${Q'}_\text{ee}^{yyyy}=0$\\ \hline
    		${\chi'}_\text{mm}^{zyy}=0$ & ${\chi'}_\text{ee}^{yyz}=0$\\ \hline
    		${\chi'}_\text{mm}^{yyz}=0$ & ${\chi'}_\text{ee}^{zyy}=0$\\ \hline
    		${S'}_\text{mm}^{yyyy}=0$ & $\chi_\text{ee}^{zz}=0$\\ \hline
    	\end{tabular}
    \end{table}

\section{Structure Synthesis and Characterization}\label{sec:structure}

\subsection{General Requirements}
The metasurface design must satisfy the conditions established in Sec.~\ref{sec:multipole_theory} to provide the desired PMC response. We begin with the case $\phi=0$ and address the case $\phi=\pi/2$ at the end of the synthesis procedure.

\begin{enumerate}
\item \textbf{Longitudinal Asymmetry}\label{req:longitudinal} \\
Equations~\eqref{eq:dipolar_pred_f} and~\eqref{eq:dipolar_pred_b} show that the PMC response must exhibit \emph{dipolar asymmetry} along the $z$ direction, with magnetic--dipole behavior under forward incidence $(+\hat{\bm{z}})$ and electric--dipole behavior under backward incidence $(-\hat{\bm{z}})$.

\item \textbf{Transverse Symmetry}\label{req:transverse}\\
The susceptibility relations in Tab.~\ref{tab:mag_group_susc} imply unit-cell nongyrotropy, and transverse reflection symmetry with respect to the $xz$ and $yz$ planes, so that a wave incident at an elevation angle $\theta$ experiences the same reflection as that at the corresponding opposite angle $-\theta$, ensuring a symmetric transverse response.

\item \textbf{Dominant Multipole Suppression}\label{req:multipole} \\
According to the multipolar decomposition hierarchy established in~\cite{jackson2021classical,raab2005multipole}, the relative magnitudes of the different multipoles follow the order
\begin{equation*}
    \scalebox{0.8}{$
    \text{electric dipole}\gg
    \begin{matrix}
        \text{electric quadrupole}\\
        \text{magnetic dipole}
    \end{matrix}\gg
    \begin{matrix}
        \text{electric octopole}\\
        \text{magnetic quadrupole}
    \end{matrix}\gg \dots$}
\end{equation*}
Therefore, among the susceptibilities to suppress according to Tab.~\ref{tab:mag_group_susc}, the most critical are the normal magnetic dipole $m_z$ for s-pol, and the perpendicular electric dipole $p_z$ together with linear electric quadrupole $q_{xx}$ for p-pol, while electric octopoles (not included in our analysis) and magnetic quadrupoles play a negligible role.
\end{enumerate}

\subsection{Metasurface Construction}

\pati{Metallization Platform and Geometry Choice}
To fulfill the longitudinal asymmetry of Requirement~\ref{req:longitudinal}), \textbf{two metallic layers} with different patterns along the $z$ axis must be employed, enabling the independent excitation of electric \emph{and} magnetic multipoles on opposite sides of the metasurface. Simultaneously, satisfying the transverse symmetry of Requirement~\ref{req:transverse}) points to the adoption of a \textbf{cross-potent particle composed of orthogonal dog-bone elements}, which provides a controlled level of cross-polarization coupling and supports a polarization-independent electromagnetic behavior~\cite{achouri2021electromagnetic,lavigne2021generalized}, while preserving mirror symmetry in the $xz$ and $yz$ planes. The orthogonal dog-bone elements will have to be designed with different sizes to suppress of parasitic multipoles that would otherwise hinder the transverse dipolar response required by Requirement~\ref{req:multipole}).

\begin{figure*}[t]
    \centering
    \includegraphics[width=\textwidth]{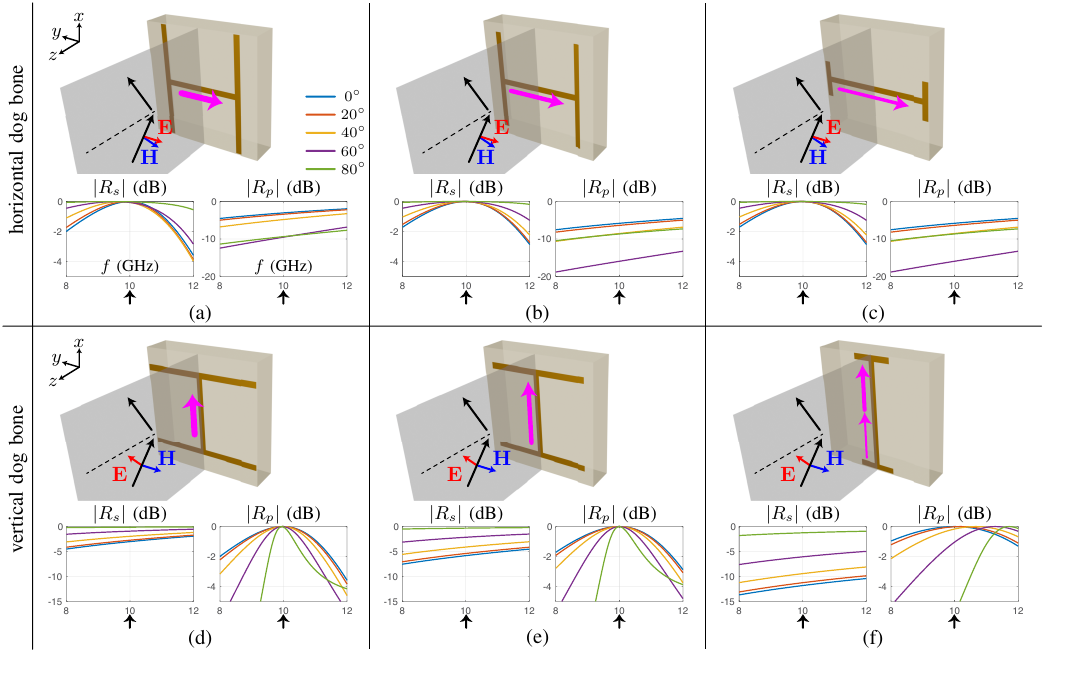}
    \vspace{-8mm}\caption{Different dog-bone frequency selective surfaces (FSSs) resonating at $f_0=10$~GHz with the reflection phase $\varphi=\pm180^\circ$ under normal incidence and related induced electric currents (magenta arrows) for oblique incidence. (a--c)~Horizontal dog-bone FSSs, sensitive to s-pol ($E_y$), arranged from smallest (a) to largest (c) dog bones. (d--f)~Same configurations as (a--c), but with vertical dog-bone orientation, sensitive to p-pol ($E_x,E_z$). The arrow widths encode the strength of the induced current. The insets show the reflection coefficients for different elevation angles, $\theta$.}
    \label{fig:FSS_dog_bone_electric_response}
\end{figure*}

\pati{Dog-Bone Electric Dipole Layer}
Let us consider then the electric-dipole layer required from the \textbf{backward side} by examining the responses of the corresponding orthogonal single dog-bone elements plotted in Fig.~\ref{fig:FSS_dog_bone_electric_response}, assuming a resonance frequency of $f_0 = 10$~GHz. The average tip-to-tip resonance length is $\ell = \lambda_0 / 4.2$\footnote{An isolated thin metallic rod in free space resonates at $\ell = \lambda_0/2$, but this resonance shifts under substrate loading, $\epsilon_\text{r,sub}=3$, and array coupling~\cite{munk2005frequency}.}. The horizontal configuration primarily responds to s-pol, while the vertical one responds to p-pol. Both geometries inherently suppress the normal magnetic dipole $m_z$ for s-pol (due to the absence of closed current loops) and the perpendicular electric dipole $p_z$ for p-pol (due to the lack of longitudinal metallization). For the horizontal dog bones [Fig.~\ref{fig:FSS_dog_bone_electric_response}(a–c)], the resonance remains nearly angle-independent under s-pol. In contrast, for vertical dog bones [Fig.~\ref{fig:FSS_dog_bone_electric_response}(d–f)], the angular response deteriorates in the largest geometry due to the excitation of the parasitic linear electric quadrupole $q_{xx}$ through inter-element coupling and  local-field gradient $\partial_x E_x$. These results reveal that \textbf{different dog-bone length along the $x$ and $y$ directions} are required for both s- and p-polarized excitations to suppress undesired multipoles. The subsequent paradox of undesired $\phi$ asymmetry will be addressed later.

\begin{figure}[H]
        \centering
        \includegraphics[width=1\columnwidth]{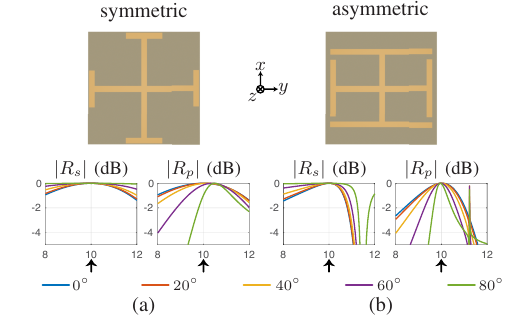}
        \vspace{-8mm}\caption{Different cross-potent FSSs resonating at $f_0=10$~GHz with the reflection phase $\varphi=\pm180^\circ$ under normal incidence s-pol and p-pol for the backward-side metallic layer. (a)~Symmetric. (b)~Asymmetric. The insets show the reflection coefficients for different angles.}
    	\label{fig:FSS_cross_potent_electric_response}
    \end{figure}

\pati{Cross-Potent Electric Response}
To achieve a polarization-independent electric response in the \textbf{backward direction}, two orthogonal dog-bone elements of different sizes may be combined into a \textbf{cross-potent configuration}, as shown in Fig.~\ref{fig:FSS_cross_potent_electric_response}. The scattering parameters in Fig.~\ref{fig:FSS_cross_potent_electric_response}(a) confirm that a symmetric configuration fail to produce angle-independent reflection, while Fig.~\ref{fig:FSS_cross_potent_electric_response}(b) confirms that, despite the \emph{geometric anisotropy}---between the $\phi=0^\circ$ ($xz$) and $\phi=\pi/2$ ($yz$) planes---the asymmetric structure provides \emph{isotropic reflection}. Specifically, shortening the vertical branch suppresses the parasitic electric quadrupole $q_{xx}$ while maintaining an electric--dipolar response over varying elevation angles.

\pati{Design of the Magnetic Dipole Layer}
To realize the \textbf{forward-side} magnetic PMC-like behavior, we adopt the asymmetric single cross-potent layer as for the back layer and use a \textbf{smaller cross-potent element as a phasing structure}. This configuration must be tailored to support anti-parallel surface currents at $f_0 = 10.6\ \text{GHz}$\footnote{The single-layer cross-potent structure served as an initial guess for the two-layer configuration, leading to a slight resonance frequency shift.} in order to generate the required magnetic dipole moments $m_{x}$ and $m_{z}$.

\begin{figure}[H]
    \centerline{\includegraphics[width=\columnwidth]{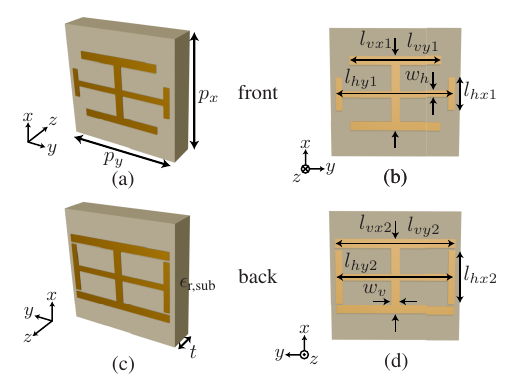}}
    \caption{Complete unit cell of the proposed PMC metasurface, composed of two metallic layers with different cross-potent structures. (a)~Perspective front view. (b)~Front view. (c)~Perspective back view. (d)~Back view. Although the structure is geometrically asymmetric in $\phi$, particularly between the planes $\phi=0^\circ$ ($xz$) and $\phi=90^\circ$ ($yz$), it will be shown to be \emph{electromagnetically symmetric} between these planes.}
    \label{fig:AMC_Structure}
    \end{figure}

\pati{Complete PMC Structure and Design Parameters}
The complete dual-layer cross-potent PMC metasurface\footnote{The dual-layer cross-potent structure is designed slightly smaller than its single-layer counterpart to mitigate the induced linear quadrupole $q_{xx}$ (see Fig.~\ref{fig:multipoles_terms_current}) which is enhanced by multiple scattering between the two metallic interfaces.} is shown in Fig.~\ref{fig:AMC_Structure}. The front layer (magnetic--dipole response) and back layer (electric--dipole response) together implement the asymmetrical excitation conditions derived from theory. The detailed geometrical parameters are given in Tab.~\ref{tab:PMC_geometrical_dimensions}, assuming a substrate permittivity $\epsilon_{\text{r,sub}} = 3$, unit cell period $p_x = p_y = 6$ mm and substrate thickness $t = 1.52$ mm.

    \begin{table}[H]
	\centering
	\setlength{\tabcolsep}{10 pt}
	\caption{Geometrical dimensions (in \textnormal{mm}) of the optimized cross-potent unit cell in Fig.~\ref{fig:AMC_Structure} with $\epsilon_\text{\textnormal{r,sub}}=3$, $p_x=p_y=6$ \textnormal{mm} and $t=1.52$ \textnormal{mm}.}
	\begin{tabular}{lcccccc}
		\hline
		\multicolumn{1}{c}{} & \multicolumn{1}{c}{$l_{\text{h}x}$} & $l_{\text{h}y}$ & $w_{\text{h}}$ & $l_{\text{v}x}$ & $l_{\text{v}y}$ & $w_{\text{v}}$ \\ \hline
		front layer & 1.53    & 5.5 & 0.3 & 3.5 & 4.2    & 0.4\\ \hline
		back layer & 2.5  & 5.5 & 0.3 & 3.5 & 5.5  & 0.4 \\ \hline
	\end{tabular}
    \label{tab:PMC_geometrical_dimensions}
\end{table}

\subsection{Asymmetry Paradox}

\pati{Enunciation}
Although geometrically asymmetric in $\phi$, the structure in Fig.~\ref{fig:AMC_Structure} exhibits an azimuthally symmetric electromagnetic response at its resonance frequency---an apparent paradox---since an isotropic response typically requires a rotationally symmetric geometry. The origin of this effect can be understood by examining the induced magnetic dipole moment generated by the localized electric current distribution, defined as~\cite{jackson2021classical}
    \begin{subequations}
    \begin{equation}\label{eq:magnetic_moment_loc_current}
        \ve{m}=\frac{1}{2}\iiint_V\ve{r}'\times\ve{J}(\ve{r}')d^3\ve{r}'.
    \end{equation}
At resonance, assuming spatially uniform and antiparallel surface currents on the two metallic layers for a magnetic dipolar response\footnote{The current distribution in Eq.~\eqref{eq:magnetic_moment_loc_current} is modeled as $\ve{J}_\text{e}(\ve{r}) = \ve{J}_\text{e0}^\text{front}(x,y)\delta(z+\Delta z/2) + \ve{J}_\text{e0}^\text{back}(x,y)\delta(z-\Delta z/2)$. Assuming uniformity and antiparallel currents, $\ve{J}_\text{e0}^\text{front} = -\ve{J}_\text{e0}^\text{back} = \ve{J}_\text{e0}^\text{micro}$.}, the transverse magnetic dipole components reduce to
\begin{align}
        m_x&=J_{\text{e}0y}^\text{micro}A_y\Delta z/2,\\
        m_y&=-J_{\text{e}0x}^\text{micro}A_x\Delta z/2,
    \end{align}
    \end{subequations}
where $A_y = w_h l_{hy}$ and $A_x = w_v l_{vx}$ denote the horizontal and vertical arm areas of the cross-potent element, and $\Delta z = t$ is the interlayer separation. When the products of arm area and induced current are comparable, the transverse magnetic dipole components ($m_x$, $m_y$) balance each other, producing an effectively isotropic magnetic response despite the anisotropic geometry.

\begin{figure}[H]
\centerline{\includegraphics[width=\columnwidth]{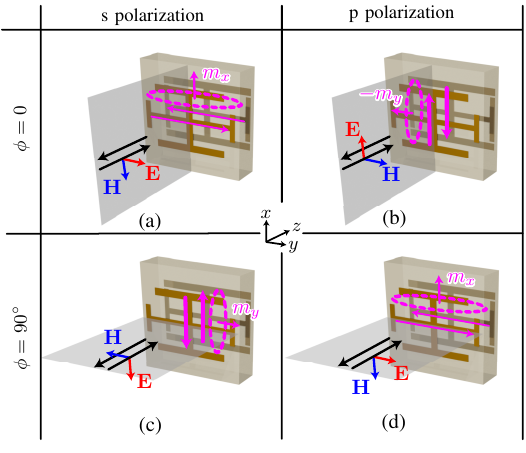}}
\caption{Azimuthal-angle electromagnetic response at resonance $f_0=10.6$~GHz for the s- and p-pol under normal illumination, $\theta=0$. (a)~s-pol at $\phi=0$. (b)~p-pol at $\phi=0^\circ$. (c)~s-pol at $\phi=\pi/2$. (d)~p-pol at $\phi=\pi/2$.}
\label{fig:phi_EM_response_normal_inc}
\end{figure}

\pati{Visual Confirmation}
This behavior is illustrated in Fig.~\ref{fig:phi_EM_response_normal_inc}. Under normal incidence, at the resonance frequency $f_0 = 10.6$~GHz, rotating along the azimuthal angle $\phi$ does not alter the electromagnetic response. 
Specifically, Fig.~\ref{fig:phi_EM_response_normal_inc}(a) and (d), as well as Fig.~\ref{fig:phi_EM_response_normal_inc}(b) and (c), exhibit identical dipole orientations and magnitudes, confirming the apparent isotropy at resonance. 
This invariance arises because the antiparallel surface currents on the two metallic layers generate balanced magnetic dipole components $m_x$ and $m_y$, whose combined radiation pattern remains independent of the azimuthal angle. 
Therefore, the observed isotropy originates not from geometric symmetry but from balanced dipolar excitation, resolving the apparent asymmetry paradox. 

\section{Demonstration}\label{sec:demonstration}
\begin{figure}[H]
    \centerline{\includegraphics[width=\columnwidth]{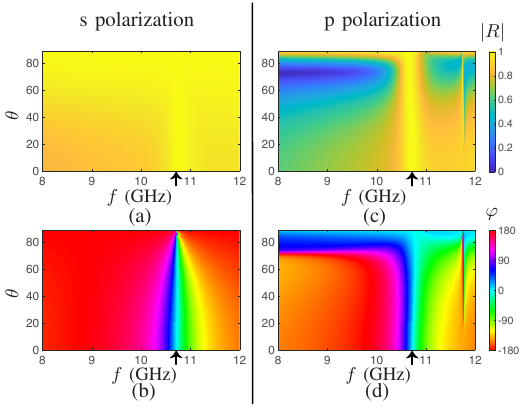}}
    \caption{Full-wave simulated reflection coefficient, $R$, versus frequency ($f$) and illumination angle ($\theta$) for the PMC structure in Fig.~\ref{fig:AMC_Structure} with the parameters in Tab.~\ref{tab:PMC_geometrical_dimensions}. (a)~Amplitude of the reflection coefficient s-pol. (b)~Phase of the reflection coefficient for s-pol. (c)~As (a) but for p-pol. (d)~Idem as (b) but for p-pol.}
    \label{fig:Full-wave}
    \end{figure}

\pati{Full-Wave Scattering Parameters}
Figure~\ref{fig:Full-wave} presents the simulated angular scattering response of the proposed PMC metasurface for both s-pol and p-pol. At the PMC resonance frequency, $f_0 = 10.6$ GHz, the structure exhibits, according to design, complete reflection with uniform phase across all elevation angles.

\begin{figure}[H]
    \centerline{\includegraphics[width=\columnwidth]{{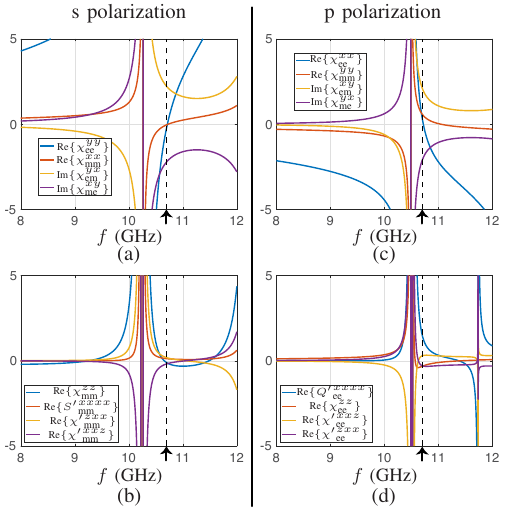}}}
    \caption{Bianisotropic susceptibilities extracted from the scattering parameters in Fig.~\ref{fig:Full-wave} (see Appendix~\ref{appendix:Susc_Extract}). (a)~Tangential dipolar susceptibilities extracted with the specification angle $\theta_\text{spec}=0^\circ$ for s-pol. (b)~Higher order susceptibilities extracted with $\theta_\text{spec}=0^\circ$ and $\theta_\text{spec}=60^\circ$ for s-pol. (c)~Idem as (a) but for p-pol. (d)~Idem as (b) but for p-pol.}
    \label{fig:extracted_susceptibility}
    \end{figure}

\pati{Consistency with Theoretical Susceptibilities}
Figure~\ref{fig:extracted_susceptibility} plots the corresponding extracted bianisotropic susceptibilities. These results validate the theoretical PMC conditions. At resonance ($f_0 = 10.6$~GHz), the heteroanisotropic conditions derived in Table~\ref{tab:mag_group_susc} are satisfied for both polarizations, with minor discrepancies in p-pol, attributed to coupling of the unit-cell dog-bone pairs.

\begin{figure}[H]
\centerline{\includegraphics[width=\columnwidth]{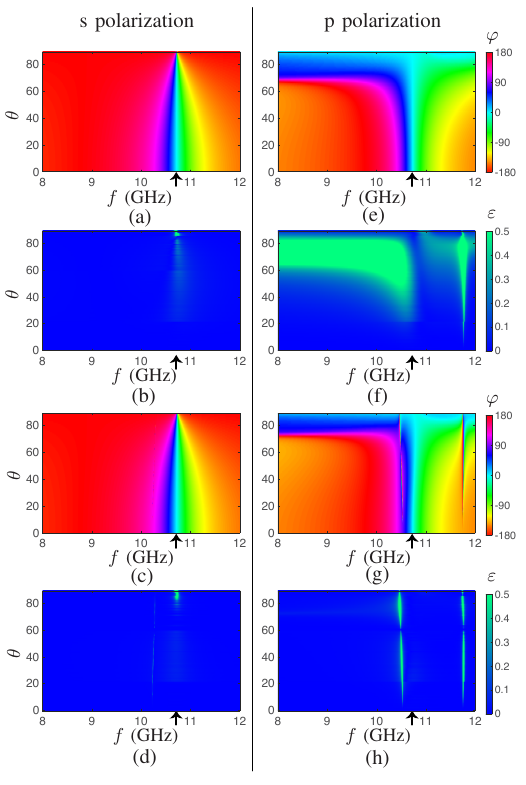}}
\caption{Theoretical (D-GSTC and DQ-GSTC) phase, $\varphi=\angle R_\text{GSTC}$, and relative residual error, $\varepsilon=|R-R_\text{GSTC}|/|R|$, for the reflection coefficient compared to the full-wave simulation in Fig.~\ref{fig:Full-wave}(c) and (d) with the specification angles $\theta_\text{spec} = 0^\circ$ and $\theta_\text{spec} = 60^\circ$. (a)~Dipolar-GSTC phase for s-pol. (b)~Dipolar-GSTC error for s-pol. (c)~DQ-GSTC phase for s-pol. (d)~DQ-GSTC error for s-pol. (e)~Idem as (a) but for p-pol. (h)~Idem as (b). (g)~Idem as (c). (h)~Idem as (d).}
\label{fig:GSTC_err}
\end{figure}

\pati{Confirmation of Elevation Angle Independency}
Figure~\ref{fig:GSTC_err} compares the theoretical scattering obtained by the D-GSTC and the DQ-GSTC models with full-wave simulations. At resonance, the D-GSTC model reproduces the angular response for both s- and p-pol, confirming the heteroanisotropic nature of the PMC metasurface. Away from resonance, however, the dipolar model captures the angular scattering for s-pol, since the magnetic quadrupole has negligible impact in this case. In contrast, for p-pol, the electric quadrupole is essential to accurately reproduce the angular dependence. The improved accuracy of the DQ-GSTC model stems from the additional poles and zeros it introduces into the metasurface transfer function, extending its validity over a limited frequency and angular range. The remaining discrepancies (sharp features) probably arise from the restrictive assumption of wavenumber-independent susceptibilities: for the considered surface, spatial dispersion due to the lattice is not properly accounted for, as the susceptibilities are taken to be incidence-angle independent. If treated explicitly, this dispersion would introduce a different set of poles and zeros. The observed sharp features therefore most likely reflect the fact that the lattice’s spatial dispersion is being approximated through quadrupolar responses, which results in spurious poles and leads to such errors.

\begin{figure}[H]
\centerline{\includegraphics[width=\columnwidth]{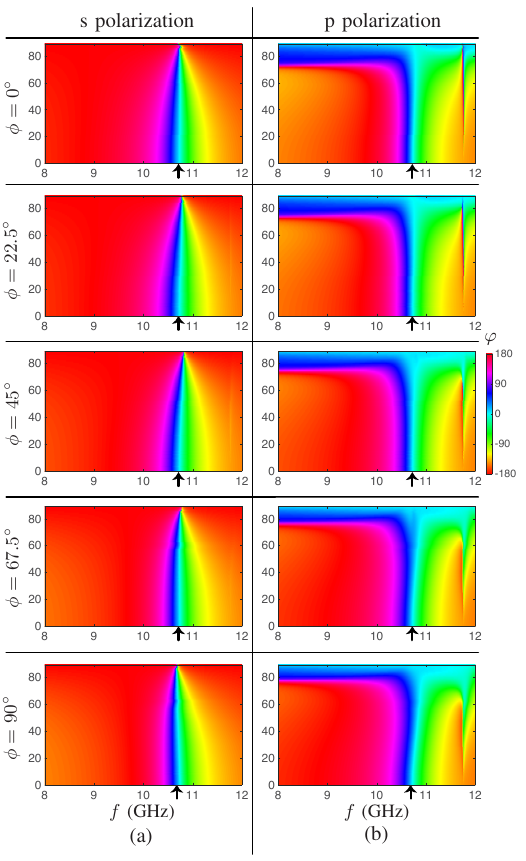}}
\caption{Full-wave simulated phase of the reflection coefficient, $\varphi$, versus frequency, $f$, and illumination angle, $\phi$, in different incidence planes for the metasurface-PMC structure in Fig.~\ref{fig:AMC_Structure}. (a)~Full-wave simulation for s-pol. (b)~Idem as (a) but for p-pol.}
\label{fig:EM_azimuth_planes}
\end{figure}

\pati{Confirmation of Quasi-Azimuth Angle Independency}
Figure~\ref{fig:EM_azimuth_planes} demonstrates the quasi-azimuthal independence of the PMC response. For both polarizations, the reflection phase remains nearly invariant with respect to the azimuthal incidence plane angle, $\phi$.

\begin{figure}[H]
\centerline{\includegraphics[width=\columnwidth]{{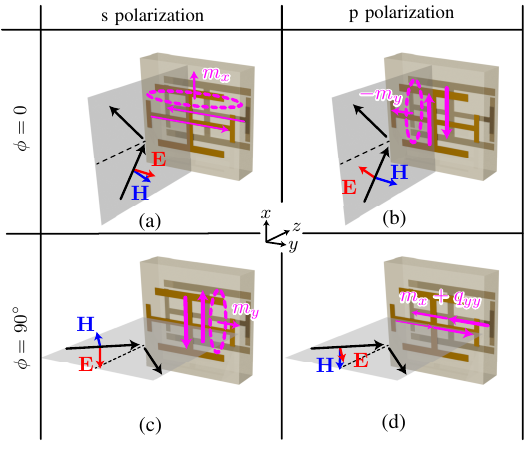}}}
\caption{Azimuthal responses under oblique illumination explaining the small variation of the PMC frequency existing for p-pol but not for s-pol in Fig.~\ref{fig:EM_azimuth_planes}. (a)~s-pol at $\phi=0$. (b)~p-pol at $\phi=0$. (c)~s-pol at $\phi=\pi/2$. (d)~p-pol at $\phi=\pi/2$.}
\label{fig:phi_EM_response_oblique_inc}
\end{figure}

\pati{Explanation of Small Discrepancy}
Figure~\ref{fig:phi_EM_response_oblique_inc} shows the emergence of a spurious linear electric quadrupole moment $q_{yy}$ under oblique incidence at $\phi = \pi/2$ for p-pol, driven by the electric field gradient $\partial_y E_y$ [see Eq.~\eqref{eq:multipole_susceptibility_based}]. While the cross-potent geometry in Fig.~\ref{fig:AMC_Structure} suppresses the parasitic quadrupole $q_{xx}$, it does not suppress $q_{yy}$, which explains the slight phase deviation observed in the full-wave results of Fig.~\ref{fig:EM_azimuth_planes} for incidence-plane angles $\phi \geq \pi/4$. This observation is consistent with the susceptibility conditions in Tab.~\ref{tab:mag_group_susc_phi_pi2}, which prescribe suppression of the linear electric quadrupole $q_{yy}$.

\section{Conclusion}\label{sec:concl}
This work has presented both a rigorous electromagnetic theory and a practical metasurface realization of a perfect magnetic conductor (PMC). A generalized theoretical framework was developed based on dipolar--quadrupolar generalized sheet transition conditions (DQ-GSTC) and multipolar surface susceptibilities, enabling accurate modeling of angle- and polarization-dependent scattering. Guided by this theory, we designed and demonstrated a physically realizable, dual layer metasurface that exhibits polarization-independent reflection and quasi-angle-independent behavior at the PMC resonance. These results establish a direct connection between theoretical PMC boundary conditions---including the essential role of electric linear quadrupole contributions---and practical metasurface implementations, paving the way for more accurate simulations, optimizations and innovative AMC designs.
\newpage

\appendices

\renewcommand{\thesection}{\Alph{section}}
\renewcommand{\thesubsection}{\thesection.\arabic{subsection}}

\setcounter{equation}{0}
\renewcommand{\theequation}{S\arabic{equation}}
\setcounter{figure}{0}
\renewcommand{\thefigure}{S\arabic{figure}}

\section{Field Specifications}\label{appendix:Fields_Specifications}
\begin{figure}[H]
\centerline{\includegraphics[width=\columnwidth]{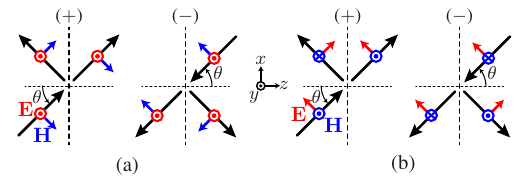}}
\vspace{-5mm}
\caption{Field specifications in the forward ($+\uve{z}$) and backward ($-\uve{z}$) incidence through a nongyrotropic metasurface, with incidence in the $xz$-plane ($\phi=0$). (a)~S-pol. (b)~P-pol.}
\label{figS:Field_Specifications}
\end{figure}

Figure~\ref{figS:Field_Specifications} shows the field specification of a uniform nongyrotropic metasurface.
The differences and averages of the fields, defined in Eqs.~\eqref{eq:DQ-GSTC} and~\eqref{eq:multipole_susceptibility_based} respectively, at $z=0$ for the two polarizations in Fig.~\ref{figS:Field_Specifications}, are, in terms of the scattering parameters, \cite{achouri2021electromagnetic}
\begin{subequations}\label{eqS:Field_Spec_SPOL}
    \begin{align}
        \Delta E_{y}&=\mp(1+R_\text{s}^\pm-T_\text{s}^\pm),\\
        \eta_0 \Delta H_{x}&=\frac{k_z}{k}(1-R_\text{s}^\pm-T_\text{s}^\pm),\\
        E_{y,\text{av}}&=(1+R_\text{s}^\pm+T_\text{s}^\pm)/2,\\
    \eta_0 H_{x,\text{av}}&=\mp\frac{k_z}{k}(1-R_\text{s}^\pm+T_\text{s}^\pm)/2,\\
    \eta_0 H_{z,\text{av}}&=\pm\frac{k_x}{k}(1+R_\text{s}^\pm+T_\text{s}^\pm)/2,
    \end{align}
\end{subequations}
for s polarization (s-pol) and
\begin{subequations}\label{eqS:Field_Spec_PPOL}
    \begin{align}
        \Delta E_{x}&=\mp\frac{k_z}{k}(1+R_\text{p}^\pm-T_\text{p}^\pm),\\
        \eta_0 \Delta H_{y}&=-(1-R_\text{p}^\pm-T_\text{p}^\pm),\\
        E_{x,\text{av}}&=\frac{k_z}{k}(1+R_\text{p}^\pm+T_\text{p}^\pm)/2\\
        E_{z,\text{av}}&=-\frac{k_x}{k}(1-R_\text{p}^\pm+T_\text{p}^\pm)/2\\
        \eta_0 H_{y,\text{av}}&=\pm(1-R_\text{p}^\pm+T_\text{p}^\pm)/2,
    \end{align}
\end{subequations}
for p polarization (p-pol). In Eqs.~\eqref{eqS:Field_Spec_SPOL} and~\eqref{eqS:Field_Spec_PPOL},  the wavenumber components are $k_x=k\sin(\theta)$ and $k_z=k\cos(\theta)$, the exponential $\text{e}^{\mp jk_xx}$ is omitted for compactness, and the upper (lower) sign corresponds to a wave propagating in the forward (backward) along the $z$-direction.

\section{Susceptibility Tensor}\label{appendix:Susceptibility_Tensor}
 
\subsection{Polarization Current Components in Terms of Multipolar Polarization Densities}
In the harmonic regime, the polarization currents~\eqref{eq:Jepol} and~\eqref{eq:Jmpol} of a multipolar density sheet in the $xy$ plane may be rewritten in component form as
\begin{subequations}\label{eqS:Je0_component_form}
	\begin{align}
        \begin{split}\label{eqS:Je0x_Mult}
    		J_{\text{e0}x}^\text{pol}&=j\omega P_x+\partial_y M_z\\
            &\quad-\frac{k^2}{2}S_{yz}-j\frac{\omega}{2}\Big(\partial_x (Q_{xx}-Q_{zz})+\partial_y Q_{xy}\Big)\\
            &\qquad-\partial_y\Big(\partial_xS_{xz}+\partial_yS_{yz}\Big),
        \end{split}\\
        \begin{split}\label{eqS:Je0y_Mult}
    		J_{\text{e0}y}^\text{pol}&=j\omega P_y-\partial_x M_z\\
            &\quad+\frac{k^2}{2}S_{xz}-j\frac{\omega}{2}\Big(\partial_x Q_{xy}+\partial_y (Q_{yy}-Q_{zz})\Big)\\
            &\qquad+\partial_x\Big(\partial_xS_{xz}+\partial_yS_{yz}\Big),
        \end{split}
	\end{align}
\end{subequations}
and
\begin{subequations}\label{eqS:Jm0_component_form}
	\begin{align}
        \begin{split}\label{eqS:Jm0x_Mult}
    		J_{\text{m0}x}^\text{pol}&=j\omega\mu_0 M_x-\frac{1}{\epsilon_0}\partial_y P_z\\
            &\quad+\frac{k^2}{2\epsilon_0}Q_{yz}-j\frac{\omega\mu_0}{2}\Big(\partial_x (S_{xx}-S_{zz})+\partial_y S_{xy}\Big)\\
            &\qquad+\frac{1}{\epsilon_0}\partial_y\Big(\partial_xQ_{xz}+\partial_yQ_{yz}\Big),
        \end{split}\\
        \begin{split}\label{eqS:Jm0y_Mult}
        J_{\text{m0}y}^\text{pol}&=j\omega \mu_0 M_y+\frac{1}{\epsilon_0}\partial_x P_z\\
            &\quad-\frac{k^2}{2\epsilon_0}Q_{xz}-j\frac{\omega\mu_0}{2}\Big(\partial_x S_{xy}+\partial_y (S_{yy}-S_{zz})\Big)\\
            &\qquad-\frac{1}{\epsilon_0}\partial_x\Big(\partial_xQ_{xz}+\partial_yQ_{yz}\Big),
        \end{split}
	\end{align}
\end{subequations}
where we used the identity $\te{a}^T=\te{a}$, since the rank-2 quadrupolar polarization moment density is a symmetric tensor by definition.

\subsection{Susceptibility Tensor Depletion}
\noindent
In this subsection, we begin with the full susceptibility tensor Eq.~\eqref{eq:susceptibility_compact_form}, which initially contains 576 components. This number is then systematically reduced by applying intrinsic tensor symmetries and the specific properties of the structure under study. For compactness, we present the derivation explicitly only for the s-pol case, noting that the procedure is analogous for the p-pol case.

\subsubsection{Dipolar Spatial-Dispersion Symmetry}\hfill \\
\textbf{First-order dipolar rank-3 tensors:} these tensors exhibit a permutation symmetry between their last two indices, namely
\begin{equation*}
    \chi'_{\text{ab},ijk} = \chi'_{\text{ab},ikj}, 
    \qquad \text{ab} \in \{\text{ee, em, me, mm}\},
\end{equation*}
which originates from their construction through the spatial-dispersion convolution integral~\cite{landau2013electrodynamics,achouri2021extension,achouri2022multipolar}. 

Without symmetry considerations, each rank-3 tensor contains $3^3=27$ components, leading to $4\times 27=108$ dipolar rank-3 susceptibilities. By imposing the permutation symmetry, the number of independent components reduces to 
\[
3\times\frac{3(3+1)}{2} = 18 \quad \text{per tensor},
\] 
hence $4\times 18=72$.  

Consequently, the total number of susceptibilities in the initial tensor decreases as 
\[
576 - (108-72) = 576 - 36 = 540,
\] 
consistent with the expected reduction.

\subsubsection{Quadrupolar Spatial-Dispersion Symmetry}\hfill \\
\textbf{Zeroth-order quadrupolar rank-3 tensors:} these tensors exhibit a permutation symmetry between their first two indices,
    \[
        Q_{\text{eb},ilj}=Q_{\text{eb},lij}, 
        \qquad 
        S_{\text{mb},ilj}=S_{\text{mb},lij}, 
        \qquad \text{b}=\{\text{e,m}\},
    \]
    which follows from the quadrupole moment tensor symmetry 
    \(\overline{\overline{Q}}_{il}=\overline{\overline{Q}}_{li}\) and 
    \(\overline{\overline{S}}_{il}=\overline{\overline{S}}_{li}\)~\cite{papas2014theory,achouri2022multipolar}.  
    This reduces the number of independent susceptibilities as
    \[
    540 - (4\times 3^2) = 504.
    \]

\textbf{First-order quadrupolar rank-4 tensors:} these tensors possess two independent permutation symmetries,
    \[
        Q'_{\text{eb},iljk}=Q'_{\text{eb},lijk}=Q'_{\text{eb},ilkj}, 
        \quad
        S'_{\text{mb},iljk}=S'_{\text{mb},lijk}=S'_{\text{mb},ilkj},
    \]
    arising from the quadrupole moment tensor symmetry (\(i \leftrightarrow l\)) and the first-order spatial-dispersion symmetry (\(j \leftrightarrow k\)).  
    The resulting reduction in independent components is
    \[
    504 - 180 = 324.
    \]

\subsubsection{Subwavelength-Thin Metasurface}\hfill \\
In the subwavelength-thin metasurface approximation, no moment density variation occurs along the normal direction of the sheet. Consequently, all susceptibility components associated with a normal derivative $\partial_z$ are suppressed. These include
    \[
    \begin{gathered}
        {\chi'}_{izj} \quad (4\times 9 = 36),\\
        Q'_{ilzj} \quad (4\times 12 = 48), 
        \qquad 
        S'_{ilzj} \quad (4\times 18 = 72),
    \end{gathered}
    \]
    which amount to $36+48+72=132$ excluded components.  

    After this suppression, the number of remaining independent susceptibilities is
    \[
    324 - 132 = 192.
    \]

\subsubsection{Incidence-Plane Restriction}\hfill \\
Given the assumption of nongyrotropy, the four GSTCs in Eq.~\eqref{eq:DQ-GSTC} decouple into two independent pairs of equations, one for each polarization. For the s-pol in the incidence plane ($\phi=0$) of Fig.~\ref{figS:Field_Specifications}, the relevant pair of GSTCs is
    \begin{subequations}\label{eqS:GSTC_SPOL}
        \begin{align}
            \Delta H_x &= J_\text{e0,y},\\
            -\Delta E_y &= -J_\text{m0,x},
        \end{align}
    \end{subequations}
where the polarization current components from~\eqref{eqS:Je0y_Mult} and~\eqref{eqS:Jm0x_Mult} become
\begin{subequations}\label{eqS:Je0Jm0_SPOL_xz-plane}
        \begin{align}
            J_{\text{e0}y}^\text{pol} &= j\omega P_y - \partial_x M_z 
            + \tfrac{k^2}{2}S_{xz} - j\tfrac{\omega}{2}\partial_x Q_{xy} 
            + \partial_x^2 S_{xz},\\
            J_{\text{m0}x}^\text{pol} &= j\omega\mu_0 M_x 
            + \tfrac{k^2}{2\epsilon_0}Q_{yz} 
            - j\tfrac{\omega\mu_0}{2}\partial_x (S_{xx}-S_{zz}),
        \end{align}
\end{subequations}
Since the plane wave is uniform along the $y$-direction, all terms involving the gradient $\partial_y$ vanish in Eqs.~\eqref{eqS:Je0_component_form} and~\eqref{eqS:Jm0_component_form}. As a result, the linear quadrupole polarization densities $Q_{yy}$ (for s-pol) and $S_{yy}$ (for p-pol), together with their associated susceptibilities, are excluded. For instance, the electric quadrupole $Q_{yy}$ suppresses the susceptibilities $Q_{\text{ee},yyj}$, $Q_{\text{em},yyj}$, ${Q'}_{\text{ee},yykj}$, and ${Q'}_{\text{em},yykj}$. This reduces the number of independent susceptibilities from $192$ to
\[
192 - 20 = 172.
\]

In Eq.~\eqref{eqS:Je0Jm0_SPOL_xz-plane}, considering the quadrupole moment tensor symmetry $\te{a}^T=\te{a}$, the polarization current response involves a total of 11 multipolar contributions: three dipole moments, $P_y$, $M_x$, $M_z$, and eight quadrupole moments, $Q_{xy}$, $Q_{yx}$, $Q_{yz}$, $Q_{zy}$, $S_{xz}$, $S_{zx}$, $S_{xx}$, and $S_{zz}$. Similarly, the GSTC for p-pol involve the complementary set of 11 polarization densities. Consequently, the final number of independent susceptibilities per polarization is
\[
\frac{172}{2} = 86.
\]

\subsubsection{Nongyrotropy}\hfill \\
According to Eq.~\eqref{eqS:Je0Jm0_SPOL_xz-plane}, and given that only the fields $E_y$, $H_x$, $H_z$, and their gradients are involved, the set of contributing susceptibilities excludes all gyrotropic terms. The multipolar polarization densities driving the s-pol response are therefore related to the fields through
\begin{equation}\label{EqS:All_Chi_SPOL}
    	\hspace{0cm}\begin{aligned}
    		&\begin{pmatrix} 
                P_y & M_x & M_z & Q_{xy} & Q_{yz} & S_{xz} & S_{xx} & S_{zz}
            \end{pmatrix}^T \\[4pt]
    		&=
            \scalebox{0.9}{$
    		\begin{pmatrix}
    			\epsilon_0 \chi_{\text{ee}}^{yy} &  \tfrac{1}{c}\chi_{\text{em}}^{yx} & \tfrac{1}{c}\chi_{\text{em}}^{yz} & \tfrac{\epsilon_0}{2k}{\chi'}_{\text{ee}}^{yxy} & \tfrac{1}{2ck}{\chi'}_{\text{em}}^{yxx} & \tfrac{1}{2ck}{\chi'}_{\text{em}}^{yxz}\\
    			\tfrac{1}{\eta_0}\chi_{\text{me}}^{xy} &  \chi_{\text{mm}}^{xx} &  \chi_{\text{mm}}^{xz} &  \tfrac{1}{2\eta_0 k}{\chi'}_{\text{me}}^{xxy} & \tfrac{1}{2k}{\chi'}_{\text{mm}}^{xxx} & \tfrac{1}{2k}{\chi'}_{\text{mm}}^{xxz}\\
    			\tfrac{1}{\eta_0}\chi_{\text{me}}^{zy} &  \chi_{\text{mm}}^{zx} &  \chi_{\text{mm}}^{zz} &  \tfrac{1}{2\eta_0 k}{\chi'}_{\text{me}}^{zxy} & \tfrac{1}{2k}{\chi'}_{\text{mm}}^{zxx} & \tfrac{1}{2k}{\chi'}_{\text{mm}}^{zxz}\\
    			\tfrac{\epsilon_0}{k}Q_{\text{ee}}^{xyy} &  \tfrac{1}{ck}Q_{\text{em}}^{xyx} & \tfrac{1}{ck}Q_{\text{em}}^{xyz} & \tfrac{\epsilon_0}{2k^2}{Q'}_{\text{ee}}^{xyxy} & \tfrac{1}{2ck^2}{Q'}_{\text{em}}^{xyxx} &  \tfrac{1}{2ck^2}{Q'}_{\text{em}}^{xyxz}\\
    			\tfrac{\epsilon_0}{k}Q_{\text{ee}}^{yzy} &  \tfrac{1}{ck}Q_{\text{em}}^{yzx} & \tfrac{1}{ck}Q_{\text{em}}^{yzz} & \tfrac{\epsilon_0}{2k^2}{Q'}_{\text{ee}}^{yzxy} & \tfrac{1}{2ck^2}{Q'}_{\text{em}}^{yzxx} & \tfrac{1}{2ck^2}{Q'}_{\text{em}}^{yzxz}\\
                \tfrac{1}{\eta_0 k}S_{\text{me}}^{xzy} &  \tfrac{1}{k}S_{\text{mm}}^{xzx} & \tfrac{1}{k}S_{\text{mm}}^{xzz} & \tfrac{1}{2\eta_0 k^2}{S'}_{\text{me}}^{xzxy} & \tfrac{1}{2k^2}{S'}_{\text{mm}}^{xzxx} &  \tfrac{1}{2k^2}{S'}_{\text{mm}}^{xzxz}\\
                \tfrac{1}{\eta_0 k}S_{\text{me}}^{xxy} & \tfrac{1}{k}S_{\text{mm}}^{xxx} & \tfrac{1}{k}S_{\text{mm}}^{xxz} & \tfrac{1}{2\eta_0 k^2}{S'}_{\text{me}}^{xxxy} & \tfrac{1}{2k^2}{S'}_{\text{mm}}^{xxxx} &  \tfrac{1}{2k^2}{S'}_{\text{mm}}^{xxxz}\\
                \tfrac{1}{\eta_0 k}S_{\text{me}}^{zzy} & \tfrac{1}{k}S_{\text{mm}}^{zzx} & \tfrac{1}{k}S_{\text{mm}}^{zzz} & \tfrac{1}{2\eta_0 k^2}{S'}_{\text{me}}^{zzxy} & \tfrac{1}{2k^2}{S'}_{\text{mm}}^{zzxx} &  \tfrac{1}{2k^2}{S'}_{\text{mm}}^{zzxz}
    		\end{pmatrix}$} \\[6pt]
    		&\qquad \cdot 
            \begin{pmatrix} 
                E_{\text{av},y} & H_{\text{av},x} & H_{\text{av},z} & \partial_x E_{\text{av},y} & \partial_x H_{\text{av},x} & \partial_x H_{\text{av},z}
            \end{pmatrix}^T.
    	\end{aligned}
    \end{equation}

This expression explicitly excludes gyrotropic susceptibilities, i.e., those coupling the s-pol multipolar densities in Eq.~\eqref{eqS:Je0Jm0_SPOL_xz-plane} to fields of the p-pol. As a result, the total number of independent susceptibilities per polarization is reduced as
\[
86 - 38 = 48.
\]

\subsubsection{Geometrical Symmetry}\hfill \\
Substituting the forward field averages from Eq.~\eqref{eqS:Field_Spec_SPOL} into Eqs.~\eqref{EqS:All_Chi_SPOL} and~\eqref{eqS:Je0Jm0_SPOL_xz-plane} yields the S-parameter form of the polarization currents\footnote{The $(+)$ superscripts, denoting forward-propagating fields, have been omitted for clarity.}:
\begin{subequations}
    \begin{align}
        \chi_A&=jk\chi_\text{ee}^{yy}+\tfrac{1}{2}kS_\text{me}^{xzy},\\
        \chi_B&=jk\chi_\text{em}^{yz}+jk\chi_\text{me}^{zy}+\tfrac{1}{2}k{\chi'}_\text{ee}^{yxy}-\tfrac{1}{2}kQ_\text{ee}^{xyy}\\
             &+\tfrac{1}{2}kS_\text{mm}^{xzz}-j\tfrac{1}{4}k{S'}_\text{me}^{xzxy},\nonumber\\
        \chi_C&=jk\chi_\text{mm}^{zz} + \tfrac{1}{2}k{\chi'}_\text{em}^{yxz} + \tfrac{1}{2}k{\chi'}_\text{mm}^{zxy} - \tfrac{1}{2}kQ_\text{em}^{xyz}\\
            &+j\tfrac{1}{4}k{Q'}_\text{ee}^{xxyxy} - k S_\text{me}^{xzy} - j\tfrac{1}{4}k{S'}_\text{mm}^{xzxz},\nonumber\\
        \chi_D&=\tfrac{1}{2}k{\chi'}_\text{mm}^{zxz}+j\tfrac{1}{4}k{Q'}_\text{em}^{xyxz}-kS_\text{mm}^{xzz}+j\tfrac{1}{2}k{S'}_\text{me}^{xzxy},\\
        \chi_E&=j\tfrac{1}{2}k{S'}_\text{mm}^{xzxz},\\
        \chi_F&=-jk\chi_\text{em}^{yx}-\tfrac{1}{2}kS_\text{mm}^{xzx},\\
        \chi_G&=-jk\chi_\text{mm}^{zx}-\tfrac{1}{2}k{\chi'}_\text{em}^{yxx}+\tfrac{1}{2}kQ_\text{em}^{xyx}+j\tfrac{1}{4}k{S'}_\text{mm}^{xzxx},\\
        \chi_H&=-\tfrac{1}{2}{\chi'}_\text{mm}^{zxx}+kS_\text{mm}^{xzx}-j\tfrac{1}{4}k{Q'}_\text{em}^{xyxx},\\
        \chi_I&=-j\tfrac{1}{2}k{S'}_\text{mm}^{xzxx},\\
        \chi_J&=jk\chi_\text{me}^{xy}+\tfrac{1}{2}kQ_\text{ee}^{yzy},\\
        \chi_K&=jk\chi_\text{mm}^{xz}+\tfrac{1}{2}k{\chi'}_\text{me}^{xxy}+\tfrac{1}{2}kQ_\text{em}^{yzz}-j\tfrac{1}{4}k{Q'}_\text{ee}^{yzxy}\\
        &-\tfrac{1}{2}kS_\text{me}^{xxy}+\tfrac{1}{2}kS_\text{me}^{zzy},\nonumber\\
        \chi_L&=\tfrac{1}{2}k{\chi'}_\text{mm}^{xxz}-j\tfrac{1}{4}k{Q'}_\text{em}^{yzxz}-\tfrac{1}{2}kS_\text{mm}^{xxz}+\tfrac{1}{2}kS_\text{mm}^{zzz}\\
        &+j\tfrac{1}{4}k{S'}_\text{me}^{xxxy}-j\tfrac{1}{4}k{S'}_\text{me}^{zzxy},\nonumber\\
        \chi_M&=j\tfrac{1}{4}k{S'}_\text{mm}^{xxxz}-j\tfrac{1}{4}k{S'}_\text{mm}^{zzxz},\\
        \chi_N&=-jk\chi_\text{mm}^{xx}-\tfrac{1}{2}kQ_\text{em}^{yzx},\\
        \chi_O&=-\tfrac{1}{2}k\chi_\text{mm}^{xxx}-j\tfrac{1}{4}{Q'}_\text{em}^{yzxx}+\tfrac{1}{2}kS_\text{mm}^{xxx}-\tfrac{1}{2}kS_\text{mm}^{zzx},\\
        \chi_P&=-j\tfrac{1}{4}k{S'}_\text{mm}^{xxxx}+j\tfrac{1}{4}k{S'}_\text{mm}^{zzxx}.
    \end{align}
\end{subequations}

If the structure has reflection symmetry with respect to the $yz$-plane, then the induced polarization currents for incidence angles $\theta$ and $-\theta$ must be identical. This condition forces the polarization currents to be even functions of $\theta$, thereby eliminating all odd-order grouped susceptibilities. The resulting expressions reduce to
	\begin{subequations}\label{eqS:Je0Jm0_SPOL_xz-plane_S_Form}
    	\begin{align}
            \eta_0J_{\text{e0}y}^\text{pol}&=\Big(\chi_A+\chi_Bs_\theta+\chi_Cs_\theta^2+\chi_Ds_\theta^3+\chi_Es_\theta^4\Big)\tfrac{1+R_\text{s}+T_\text{s}}{2}\nonumber\\
            &\quad+\Big(\chi_F+\chi_Gs_\theta+\chi_Hs_\theta^2+\chi_Is_\theta^3\Big)\tfrac{(1-R_\text{s}+T_\text{s})c_\theta}{2},\\
    		J_{\text{m0}x}^\text{pol}&=\Big(\chi_J+\chi_Ks_\theta+\chi_Ls_\theta^2+\chi_Ms_\theta^3\Big)\tfrac{1+R_\text{s}+T_\text{s}}{2}\nonumber\\
            &\quad+\Big(\chi_N+\chi_Os_\theta+\chi_Ps_\theta^2\Big)\tfrac{(1-R_\text{s}+T_\text{s})c_\theta}{2}.
    	\end{align}
    \end{subequations}
where the \emph{grouped coefficients} have been renamed to highlight the dominant multipolar, nonlocal contributions and absorbing real-valued constants for simplicity, where the absorbing terms are
\begin{subequations}
        \begin{align}
            \chi_A&=jk\chi_\text{ee}^{yy},\;\chi_C=jk\chi_\text{mm}^{zz},\;\chi_E=jk{S'}_\text{mm}^{xzxz},\\
            \chi_F&=-jk\chi_\text{em}^{yx},\;\chi_H=-{\chi'}_\text{mm}^{zxx},\;\chi_J=jk\chi_\text{me}^{xy},\\
            \chi_L&=k{\chi'}_\text{mm}^{xxz},\;\chi_N=-jk\chi_\text{mm}^{xx},\;\chi_P=-jk{S'}_\text{mm}^{xxxx}.
        \end{align}
\end{subequations}

Thus, from the previous $48$ nongyrotropic susceptibilities, reflection symmetry eliminates $26$ terms, leaving $22$ independent terms, which can be further reduced to \textbf{$9$ independent effective susceptibilities}.

\section{Well-Posed DQ-GSTC Problem}\label{appendix:Well-Posed}
As seen in the previous section, the reduced GSTC system obtained from symmetry and various restrictions is restricted to the following $9$ grouped susceptibilities:
\[
\big\{\chi_\text{ee}^{yy},\;\chi_\text{mm}^{zz},\;\chi_\text{em}^{yx},\;\chi_\text{me}^{xy},\;
\chi_\text{mm}^{xx},\;\chi_\text{mm}^{zxx},\;{\chi'}_\text{mm}^{xxz},\;
{S'}_\text{mm}^{xxxx},\;{S'}_\text{mm}^{xzxz}\big\}.
\]

Since ${S'}_\text{mm}^{xzxz}$ contributes only through a quartic angular dependence $s_\theta^4$, it is assumed negligible. The \textbf{number of unknowns is thus reduced to $8$}, which exactly matches the number of independent conditions provided by the DQ-GSTCs. The problem is therefore well posed. In this case, the s-pol GSTCs take the form
\begin{subequations}\label{eqS:DQ-GSTC_Sform_SPOL}
    	\begin{align}
            (1-R_\text{s}^{\pm}-T_\text{s}^{\pm})c_\theta
    		&=\Big(jk\chi_\text{ee}^{yy}+jk\chi_\text{mm}^{zz}s_\theta^2\Big)\tfrac{1+R_\text{s}^{\pm}+T_\text{s}^{\pm}}{2}\nonumber\\
            &\quad\pm\Big(-jk\chi_\text{em}^{yx}-k\chi_\text{mm}^{zxx}s_\theta^2\Big)\tfrac{(1-R_\text{s}^{\pm}+T_\text{s}^{\pm})c_\theta}{2},\\
            \mp(1+R_\text{s}^{\pm}-T_\text{s}^{\pm})
    		&=\Big(jk\chi_\text{me}^{xy}+k{\chi'}_\text{mm}^{xxz}s_\theta^2\Big)\tfrac{1+R_\text{s}^{\pm}+T_\text{s}^{\pm}}{2}\nonumber\\
            &\quad\pm\Big(-jk\chi_\text{mm}^{xx}-jk{S'}_\text{mm}^{xxxx}s_\theta^2\Big)\tfrac{(1-R_\text{s}^{\pm}+T_\text{s}^{\pm})c_\theta}{2}.
    	\end{align}
\end{subequations}

In summary, the initial $9$ grouped susceptibilities have been reduced to $8$ upon neglecting the quartic term ${S'}_\text{mm}^{xzxz}$, and the DQ-GSTCs provide exactly $8$ independent equations, ensuring that the problem is well posed. Similarly, for p-pol, one susceptibility (${Q'}_\text{ee}^{xzxz}$) is neglected for the same reason, which again yields a well-posed system.

\section{Scattering Parameters in Terms of Multipolar Susceptibilities}\label{appendix:Scatt_Para}
The s-pol DQ-GSTCs in Eq.~\eqref{eqS:DQ-GSTC_Sform_SPOL} yield four equations per transformation, with four unknown scattering coefficients. Two transformations are considered: one forward- and one backward-propagating plane wave at angle $\theta_{\text{spec.}}=\theta$ (see Fig.~\ref{figS:Field_Specifications}). Solving this system provides explicit expressions for the angular scattering coefficients $R_\text{s}^\pm(\theta)$ and $T_\text{s}^\pm(\theta)$ in terms of the multipolar susceptibilities. They are given by
\begin{subequations}\label{eqS:angular_scattering_SPOL}
    \begin{equation}
\begin{aligned}
    R^+_s(\theta)&=\Big[-2jk\chi_\text{ee}^{yy}-2jk\left(\chi_\text{em}^{yx}-\chi_\text{me}^{xy}\right)c_\theta\\
    &2jk\chi_\text{mm}^{xx}s_\theta^2-2jk\chi_\text{mm}^{zz}c_\theta^2+2k\left({\chi'}_\text{mm}^{xxz}-{\chi'}_\text{mm}^{zxx}\right)c_\theta s_\theta^2\\
    &+2jk{S'}_\text{mm}^{xxxx}c_\theta^2 s_\theta^2\Big]/\Delta,
\end{aligned}
\end{equation}
\begin{equation}
\begin{aligned}
    T^+_s(\theta)&=\Big[k^2\chi_\text{ee}^{yy}\chi_\text{mm}^{xx}+\left(2-jk\chi_\text{em}^{yx}\right)\left(2-jk\chi_\text{me}^{xy}\right)\\
    &+k^2\Big(\chi_\text{ee}^{yy}{S'}_\text{mm}^{xxxx}+\chi_\text{mm}^{xx}\chi_\text{mm}^{zz}+j\chi_\text{em}^{yx}{\chi'}_\text{mm}^{xxz}+j\chi_\text{me}^{xy}{\chi'}_\text{mm}^{zxx}\\
    &-2\left({\chi'}_\text{mm}^{xxz}+{\chi'}_\text{mm}^{zxx}\right)\Big)s_\theta^2+\\
    &+k^2\left(\chi_\text{mm}^{zz}{S'}_\text{mm}^{xxxx}+{\chi'}_\text{mm}^{xxz}{\chi'}_\text{mm}^{zxx}\right)s_\theta^4\Big]c_\theta/\Delta,
\end{aligned}
\end{equation}
\begin{equation}
\begin{aligned}
    R^-_s(\theta)&=\Big[-2jk\chi_\text{ee}^{yy}+2jk\left(\chi_\text{em}^{yx}-\chi_\text{me}^{xy}\right)c_\theta\\
    &2jk\chi_\text{mm}^{xx}s_\theta^2-2jk\chi_\text{mm}^{zz}c_\theta^2-2k\left({\chi'}_\text{mm}^{xxz}-{\chi'}_\text{mm}^{zxx}\right)c_\theta s_\theta^2\\
    &+2jk{S'}_\text{mm}^{xxxx}c_\theta^2 s_\theta^2\Big]/\Delta,
\end{aligned}
\end{equation}
\begin{equation}
\begin{aligned}
    T^-_s(\theta)&=\Big[k^2\chi_\text{ee}^{yy}\chi_\text{mm}^{xx}+\left(2+jk\chi_\text{em}^{yx}\right)\left(2+jk\chi_\text{me}^{xy}\right)\\
    &+k^2\Big(\chi_\text{ee}^{yy}{S'}_\text{mm}^{xxxx}+\chi_\text{mm}^{xx}\chi_\text{mm}^{zz}+j\chi_\text{em}^{yx}{\chi'}_\text{mm}^{xxz}+j\chi_\text{me}^{xy}{\chi'}_\text{mm}^{zxx}\\
    &+2\left({\chi'}_\text{mm}^{xxz}+{\chi'}_\text{mm}^{zxx}\right)\Big)s_\theta^2+\\
    &+k^2\left(\chi_\text{mm}^{zz}{S'}_\text{mm}^{xxxx}+{\chi'}_\text{mm}^{xxz}{\chi'}_\text{mm}^{zxx}\right)s_\theta^4\Big]c_\theta/\Delta,
\end{aligned}
\end{equation}
where
\begin{equation}
\begin{aligned}
    \Delta&=2jk\chi_\text{ee}^{yy}+\left(4+k^2\chi_\text{em}^{xy}\chi_\text{me}^{xy}-k^2\chi_\text{ee}^{yy}\chi_\text{mm}^{xx}\right)c_\theta\\
    &+2jk\chi_\text{mm}^{xx}c^2_\theta+2jk\chi_\text{mm}^{zz}s^2_\theta\\
    &-k^2\left(\chi_\text{mm}^{xx}\chi_\text{mm}^{zz}+\chi_\text{ee}^{yy}{S'}_\text{mm}^{xxxx}+j\chi_\text{em}^{yx}{\chi'}_\text{mm}^{xxz}+j\chi_\text{me}^{xy}{\chi'}_\text{mm}^{zxx}\right)c_\theta s^2_\theta\\
    &+2jk{S'}_\text{mm}^{xxxx}c^2_\theta s^2_\theta-k^2\left(\chi_\text{mm}^{zz}{S'}_\text{mm}^{xxxx}+{\chi'}_\text{mm}^{xxz}{\chi'}_\text{mm}^{zxx}\right)c_\theta s^4_\theta.
\end{aligned}
\end{equation}
\end{subequations}

\section{PMC Grouped Susceptibility Condition for Elevation Angle Independence}\label{appendix:PMC_susc}
A PMC corresponds to angle-independent reflection, i.e., $R(\theta)=+1,~\forall\theta$. The angle-dependent DQ-GSTC in Eq.~\eqref{eqS:DQ-GSTC_Sform_SPOL} lead to angle-dependent coefficients $R^\pm_s(\theta)$ and $T^\pm_s(\theta)$. By setting to zero the homosusceptibilities $\chi_\text{ee}^{yy}$, $\chi_\text{mm}^{xx}$, $\chi_\text{mm}^{zz}$, ${S'}_\text{mm}^{xxxx}$, ${\chi'}_\text{mm}^{zxx}$ and ${\chi'}_\text{mm}^{xxz}$ in Eq.~\eqref{eqS:DQ-GSTC_Sform_SPOL}, the angle-dependent DQ-GSTCs reduce to
    \begin{subequations}\label{eqS:DQ-GSTCs_Sform_SPOL_ang_ind}
	\begin{align}
        \begin{split}
            (1-&R_\text{s}^{\pm}-T_\text{s}^{\pm})\\
    		&=\pm\Big(-jk\chi_\text{em}^{yx}\Big)(1-R_\text{s}^{\pm}+T_\text{s}^{\pm})/2,
        \end{split}\\
        \begin{split}
            \mp(1+&R_\text{s}^{\pm}-T_\text{s}^{\pm})\\
    		&=\Big(jk\chi_\text{me}^{xy}\Big)(1+R_\text{s}^{\pm}+T_\text{s}^{\pm})/2.
        \end{split}
	\end{align}
    \end{subequations}
Solving Eq.~\eqref{eqS:DQ-GSTCs_Sform_SPOL_ang_ind} yields
    \begin{subequations}\label{eqS:RTs_ang_ind}
    \begin{align}
        R_\text{s}^{\pm}&=\pm j2\frac{k(\chi_\text{em}^{yx}-\chi_\text{me}^{xy})}{k^2\chi_\text{em}^{yx}\chi_\text{me}^{xy}+4},\\
        T_\text{s}^{\pm}&=\frac{(2\pm jk\chi_\text{em}^{yx})(2\pm jk\chi_\text{me}^{xy})}{k^2\chi_\text{em}^{yx}\chi_\text{me}^{xy}+4}.
    \end{align}
    \end{subequations}
Imposing the reciprocity condition $\chi_\text{me}^{xy}=-\chi_\text{em}^{yx}$~\cite{achouri2021electromagnetic}, Eq.~\eqref{eqS:RTs_ang_ind} simplifies to
    \begin{subequations}\label{eqS:RTs_ang_ind_rec}
    \begin{align}
        R_\text{s}^{\pm}&=\pm j4\frac{k\chi_\text{em}^{yx}}{4-k^2{\chi_\text{em}^{yx}}^2},\label{eqS:RFBs_ang_ind_rec}\\
        T_\text{s}^{\pm}&=\frac{4+k^2{\chi_\text{em}^{yx}}^2}{4-k^2{\chi_\text{em}^{yx}}^2}.
    \end{align}
    \end{subequations}
Enforcing the PMC condition $R_\text{s}^+=+1$ gives $\chi_\text{em}^{yx}=-j2/k$ and $\chi_\text{me}^{xy}=j2/k$, corresponding to a PMC/PEC solution. Similarly, for p-pol, the condition yields $\chi_\text{em}^{xy}=j2/k$ and $\chi_\text{me}^{yx}=-j2/k$.  
The heteroanisotropic PMC solution takes the form
    \begin{equation}\label{eqS:PMC_heteroanisotropic}
        \tes{\upchi}_\text{em}=
        \begin{pmatrix}
            0 & j\frac{2}{k} \\
        -j\frac{2}{k} & 0
        \end{pmatrix}\quad \text{and}\quad \tes{\upchi}_\text{me}=
        \begin{pmatrix}
            0 & j\frac{2}{k} \\
            -j\frac{2}{k} & 0
        \end{pmatrix},
    \end{equation}
    where the surface susceptibilities represent a nonlocal, heteroanisotropic omega-type medium~\cite{achouri2021electromagnetic}.

\section{Susceptibility Extraction from Scattering Parameters}\label{appendix:Susc_Extract}

The s-pol DQ-GSTCs in Eq.~\eqref{eqS:DQ-GSTC_Sform_SPOL} yield two equations per transformation, with eight susceptibility unknowns in total. Considering four transformations---two forward- and two backward-propagating plane waves at angles $\theta_{\text{spec.},1}=0$ and $\theta_{\text{spec.},2}=\theta_0$---the extracted susceptibility equations are obtained as
\begin{subequations}
    \begin{align}
        \begin{split}
        &\chi_\text{ee}^{yy}=\\
        &\frac{2}{jk}\frac{1 - R_\text{s}^{-}(0) - R_\text{s}^{+}(0) + R_\text{s}^{-}(0) R_\text{s}^{+}(0) - T_\text{s}^{-}(0) T_\text{s}^{+}(0)}
       {1 - R_\text{s}^{-}(0) R_\text{s}^{+}(0) + T_\text{s}^{-}(0) + T_\text{s}^{+}(0) + T_\text{s}^{-}(0) T_\text{s}^{+}(0)},
    \end{split}\\
    \begin{split}
        &\chi_\text{em}^{yx}=\\
        &\frac{2}{jk}\frac{-R_\text{s}^{-}(0) + R_\text{s}^{+}(0) - T_\text{s}^{-}(0) + T_\text{s}^{+}(0)}
       {1 - R_\text{s}^{-}(0) R_\text{s}^{+}(0) + T_\text{s}^{-}(0) + T_\text{s}^{+}(0) + T_\text{s}^{-}(0) T_\text{s}^{+}(0)},
    \end{split}\\
    \begin{split}
        &\chi_\text{me}^{xy}=\\
        &\frac{2}{jk}\frac{R_\text{s}^{-}(0) - R_\text{s}^{+}(0) - T_\text{s}^{-}(0) + T_\text{s}^{+}(0)}
       {1 - R_\text{s}^{-}(0) R_\text{s}^{+}(0) + T_\text{s}^{-}(0) + T_\text{s}^{+}(0) + T_\text{s}^{-}(0) T_\text{s}^{+}(0)},
    \end{split}\\
    \begin{split}
        &\chi_\text{mm}^{xx}=\\
        &\frac{2}{jk}\frac{1 + R_\text{s}^{-}(0) + R_\text{s}^{+}(0) + R_\text{s}^{-}(0) R_\text{s}^{+}(0) - T_\text{s}^{-}(0) T_\text{s}^{+}(0)}
       {1 - R_\text{s}^{-}(0) R_\text{s}^{+}(0) + T_\text{s}^{-}(0) + T_\text{s}^{+}(0) + T_\text{s}^{-}(0) T_\text{s}^{+}(0)},
    \end{split}\\
    \begin{split}
        &\chi_\text{mm}^{zz}=\\
        &\frac{2}{jk}\text{csc}^2(\theta_0)\Bigg(-\frac{jk}{2}\chi_\text{ee}^{yy}\\
        &+\frac{-1 + R_\text{s}^{-}(\theta_0) + R_\text{s}^{+}(\theta_0) - R_\text{s}^{-}(\theta_0) R_\text{s}^{+}(\theta_0) + T_\text{s}^{-}(\theta_0) T_\text{s}^{+}(\theta_0)}
     {R_\text{s}^{-}(\theta_0) R_\text{s}^{+}(\theta_0) - \left(1 + T_\text{s}^{-}(\theta_0)\right) \left(1 + T_\text{s}^{+}(\theta_0)\right)}\cos(\theta_0)\Bigg),
    \end{split}\\
    \begin{split}
        &{\chi'}_\text{mm}^{zxx}=\\
        &\frac{2}{k}\text{csc}^2(\theta_0)\Bigg(\frac{R_\text{s}^{-}(\theta_0) - R_\text{s}^{+}(\theta_0) + T_\text{s}^{-}(\theta_0) - T_\text{s}^{+}(\theta_0)}
     {R_\text{s}^{-}(\theta_0) R_\text{s}^{+}(\theta_0) - \left(1 + T_\text{s}^{-}(\theta_0)\right)\left(1 + T_\text{s}^{+}(\theta_0)\right)}
    \\
    &+\frac{jk}{2}\chi_\text{em}^{yx}\frac{1 - R_\text{s}^{-}(\theta_0) R_\text{s}^{+}(\theta_0) + T_\text{s}^{-}(\theta_0) + T_\text{s}^{+}(\theta_0) + T_\text{s}^{-}(\theta_0) T_\text{s}^{+}(\theta_0)}
     {R_\text{s}^{-}(\theta_0) R_\text{s}^{+}(\theta_0) - \left(1 + T_\text{s}^{-}(\theta_0)\right)\left(1 + T_\text{s}^{+}(\theta_0)\right)}\Bigg),
    \end{split}\\
    \begin{split}
        &{\chi'}_\text{mm}^{xxz}=\\
        &\frac{2}{k}\text{csc}^2(\theta_0)\Bigg(\frac{-R_\text{s}^{-}(\theta_0) + R_\text{s}^{+}(\theta_0) + T_\text{s}^{-}(\theta_0) - T_\text{s}^{+}(\theta_0)}
     {R_\text{s}^{-}(\theta_0) R_\text{s}^{+}(\theta_0) - \left(1 + T_\text{s}^{-}(\theta_0)\right)\left(1 + T_\text{s}^{+}(\theta_0)\right)}
    \\
    &+\frac{jk}{2}\chi_\text{me}^{xy}\frac{1 - R_\text{s}^{-}(\theta_0) R_\text{s}^{+}(\theta_0) + T_\text{s}^{-}(\theta_0) + T_\text{s}^{+}(\theta_0) + T_\text{s}^{-}(\theta_0) T_\text{s}^{+}(\theta_0)}
     {R_\text{s}^{-}(\theta_0) R_\text{s}^{+}(\theta_0) - \left(1 + T_\text{s}^{-}(\theta_0)\right)\left(1 + T_\text{s}^{+}(\theta_0)\right)}\Bigg),
    \end{split}\\
    \begin{split}
        &{S'}_\text{mm}^{xxxx}=\\
        &\frac{2}{jk}\text{csc}^2(\theta_0)\Bigg(-\frac{jk}{2}\chi_\text{mm}^{xx}\\
        &+\frac{1 + R_\text{s}^{-}(\theta_0) + R_\text{s}^{+}(\theta_0) + R_\text{s}^{-}(\theta_0) R_\text{s}^{+}(\theta_0) - T_\text{s}^{-}(\theta_0) T_\text{s}^{+}(\theta_0)}
         {1 - R_\text{s}^{-}(\theta_0) R_\text{s}^{+}(\theta_0) + T_\text{s}^{-}(\theta_0) + T_\text{s}^{+}(\theta_0) + T_\text{s}^{-}(\theta_0) T_\text{s}^{+}(\theta_0)}
        \sec(\theta_0)\Bigg).
    \end{split}
    \end{align}
\end{subequations}
%


\bibliography{Perfect_AMC}
\bibliographystyle{ieeetr}

\end{document}